\documentclass[preprint]{aastex} 

\newcommand{\be}{\begin{equation}}
\newcommand{\ee}{\end{equation}}
\newcommand{\nd}{\nodata}

\newfont{\Bd}{msbm10 scaled\magstep1}

\begin{document}


\title{Spatially Resolved Circumstellar Structure of Herbig Ae/Be Stars 
in the Near-Infrared}

\author{Rafael Millan-Gabet\altaffilmark{1,2} 
and F. Peter Schloerb}
\affil{Department of Physics and Astronomy\\ 
	University of Massachusetts at Amherst \\
	Amherst, MA 01003}
\email{rmillan@cfa.harvard.edu\\schloerb@astro.umass.edu}

\altaffiltext{1}{present address: Harvard-Smithsonian Center for Astrophysics,
    60 Garden Street, Cambridge, MA 02138}
\altaffiltext{2}{Michelson Postdoctoral Fellow}

\and

\author{Wesley A. Traub}
\affil{Harvard-Smithsonian Center for Astrophysics \\
	Cambridge, MA 02138}
\email{wtraub@cfa.harvard.edu}

\begin{abstract}

We have conducted the first systematic study of Herbig Ae/Be stars
using the technique of long baseline stellar interferometry in the
near-infrared, with the objective of characterizing the distribution
and properties of the circumstellar dust responsible for the excess
near-infrared fluxes from these systems. The observations for this
work have been conducted at the Infrared Optical Telescope Array
(IOTA).  The principal result of this paper is that the interferometer
resolves the source of infrared excess in 11 of the 15 systems
surveyed. A new binary, MWC~361-A, has been detected
interferometrically for the first time. 

The visibility data for all the sources has been interpreted within
the context of four simple models which represent a range of plausible
representations for the brightness distribution of the source of
excess emission: a Gaussian, a narrow uniform ring, a flat blackbody
disk with a single temperature power law, and an infrared companion.
We find that the characteristic sizes of the near-infrared emitting
regions are larger than previously thought ($0.5-5.9$ AU, as given by
the FWHM of the Gaussian intensity). A further major result of this
paper is that the sizes measured, when combined with the observed
spectral energy distributions, essentially rule out accretion disk
models represented by blackbody disks with the canonical $T(r) \propto
r^{-3/4}$ law.  We also find that, within the range observed in this
study, none of the sources (except the new binary) shows varying
visibilities as the orientation of the interferometer baseline
changes. This is the expected behaviour for sources which appear
circularly symmetric on the sky, and for the sources with the largest
baseline position angle coverage (AB~Aur, MWC~1080-A) asymmetric
brightness distributions (such as inclined disks or binaries) become
highly unlikely.

Taken as an ensemble, with no clear evidence in favor of axi-symmetric
structure, the observations favor the interpretation that the
circumstellar dust is distributed in spherical envelopes (the Gaussian
model) or thin shells (the ring model). This interpretation is also
supported by the result that the measured sizes, combined with the
excess near-infrared fluxes, imply emission of finite optical depth,
as required by the fact that the central stars are optically
visible. The measured sizes and brightnesses do not correlate strongly
with the luminosity of the central star.  Moreover, in two cases, the
same excess is observed from circumstellar structures that differ in
size by more than a factor of two, and surround essentially identical
stars.  Therefore, different physical mechanisms for the near-infrared
emission may be at work in different cases, or alternatively, a single
underlying mechanism with the property that the same infrared excess
is produced on very different physical scales.
  
\end{abstract}


\section{INTRODUCTION} \label{intro}

The Herbig Ae/Be (HAEBE) stars are understood to be young stellar
objects of intermediate mass ($1.5 < M/M_{\sun} < 10$)
\citep{Herbig:1960}. Observationally, they are defined by spectral
types B to F8 with emission lines and the presence of infrared (IR)
to sub-mm excess flux due to hot and cool dust. The pre-main sequence
nature of HAEBE stars is now well established, based on comparison of
their location in the HR diagram with theoretical evolutionary tracks
\citep{Strom:1972,Cohen:1979,Ancker:1998,Palla:1991}, their 
physical association with the dark clouds in
which most HAEBE stars are found \citep{Finkenzeller:1984b}, and
their high rotational velocities \citep{Finkenzeller:1985}.
The question of the geometrical distribution of the cool material
which gives rise to the characteristic long wavelength excess,
however, has remained controversial and is still the subject of
considerable debate.

In the case of the T Tauri stars, the excess continuum flux is
successfully reproduced without extinguishing the central star at
visible wavelengths by a model in which a circumstellar (CS) disk is
heated both by the central star and by active accretion. The disk is
also the source of the observed activity (permitted and forbidden
emission line intensities and assymetric profiles, UV excess, veiling
of absorption lines) through magnetospheric infall onto the stellar
surface \citep*{Muzerolle:1998}.  An important consideration in
understanding the success of this model in this case is the fact that
the level of activity observed can not be explained by any known
mechanism involving the star alone.

For the HAEBE stars on the other hand, the situation is considerably
more confused, since the much higher stellar luminosities permit a
number of alternative physical processes to reproduce the
observations. Moreover, the much smaller number of sources and their
more rapid evolution make statistical analyses less reliable. As an
example, \citet{Bohm:1995} conclude that the origin of outflowing wind
in HAEBE stars arises in an accretion disk boundary layer as in the T
Tauri stars, whereas \citet{Corcoran:1998} favor the interpretation
that the outflow arises in the star itself.  Likewise, the
interpretation of the line profiles of forbidden lines is
controversial.  In the case of the classical TTS, nearly all forbidden
lines lack a redshifted component, a fact attributed to occultation of
the receding part of the wind by the opaque CS disk.  A similar
analysis for the HAEBE stars however shows that the forbidden lines
are symmetric and therefore inconsistent with such a model
\citep{Bohm:1994}, although \citet{Corcoran:1997} reach opposite
conclusions based on similar data.

In a much cited paper, \citet{Hillenbrand:1992} (hereafter HSVK)
classified 47 HAEBE stars in three groups according to the shape of
their spectral energy distributions (SED).  Group~I contained stars
with IR spectral distributions $\lambda F_{\lambda} \sim
\lambda^{-4/3}$, reminiscent of the spectral shape expected for a
star~-~disk system in which the disk both reprocesses starlight and/or
is self luminous via active accretion onto the central star
\citep{Basri:1993}. Group~II contained stars with flat or rising IR
spectra, which require the presence of gas and dust not confined to a
flat disk.  Finally, Group~III contained stars with small or no IR
excess, similar to the classical Be stars, in which the small excesses
above photospheric levels arises from free-free emission in a gaseous
disk or envelope.  HSVK modelled the SEDs of Group~I stars assuming
that the IR excess does in fact arise in a reprocessing and actively
accreting disk, and they proposed an evolutionary sequence Group~II
$\rightarrow$ Group~I $\rightarrow$ Group~III, where the originally
complex gas and dust environment that results from the star formation
process leads naturally to the formation of an accretion disk which
gradually disappears.  HSVK found that the SEDs of Group~I sources
could be satisfactorily fit provided that the accretion rates are
relatively high ($6 \times 10^{-7} \leq \dot{M} \leq 8 \times 10^{-5}
\, M_{\sun} \, yr^{-1}$) and that opacity holes of a few stellar radii
exists in the inner region of the disk. These opacity holes were
tentatively interpreted as being due to dust sublimation or,
alternatively, to the interaction of the stellar magnetosphere with
the disk, which lifts gas from the disk and forces accreted gas and
dust to proceed along the stellar magnetic field lines
\citep{Konigl:1991}. Although the same conclusions were reached by
\citet{Hartmann:1993}, these authors also pointed out the fact that
for the high accretion rates required to fit the SEDs, the {\it gas}
in the region interior to the inner hole would be optically thick,
resulting in additional emission which is not observed. Moreover, the
derived accretion rates are inconsistent with the observed absence of
optical veiling in the spectra of HAEBE stars \citep{Ghandour:1994}.
The magnetosphere hypothesis, on the other hand, requires stellar
magnetic fields strengths ($\sim 10$ KG) not normally observed in
early-type stars.

At the very least then, it can be said that, although it would appear
natural to assume that similar physical processes occur in the CS
environment of T Tauri and HAEBE stars, the existence of accretion
disks around HAEBE stars is far from being firmly established.  As a
consequence, alternative ideas have been considered to explain the
observed IR excess, such as symmetric dust (and gas) envelopes
\citep{Berrilli:1992,Pezzuto:1997}. It must be said that in this case
too there is lack of consensus about the precise properties of these
envelopes (density profiles, emission mechanisms), or even the
question of whether or not envelopes are capable at all of explaining
the observed SEDs (see for example \citealt{Miro:1997} and
\citealt{Natta:1993a}). One recurrent difficulty with symmetrical
envelopes is that they tend to result in higher visual extinctions
toward the central star than are observed, which has led some authors
to postulate special geometries which provide a low optical depth line
of sight to the star \citep{Hartmann:1993}, or to models of more
tenous envelopes consisting of very small, transiently heated dust
grains \citet{Hartmann:1993}.

In summary, there is currently no physically accepted model to
account for the IR emission of HAEBE stars. At the same time, the
answers to many questions on the evolutionary status and the origin of
the activity and variability depend critically on the relative
importance of CS distribution of material in disks or envelopes at
different spatial scales. A desirable approach is to constrain
the models not only with respect to the observed SEDs, but also with
respect to the spatial distribution of the IR emission. Clearly, one
way to make progress toward this important question is to obtain
observations of the near-IR emission with sufficient angular
resolution to resolve the structure around the central star, 
and that is the principal goal of the work reported here.

In $\S$~\ref{exp} we describe our experimental procedure
and data analysis method, including a novel technique for calibration
of visibility amplitude data in the presence of atmospheric tip-tilt
wavefront errors. In $\S$~\ref{sample} we present the list of HAEBE
stars observed, log of observations and the calibrated visibility
data.  In $\S$~\ref{fluxes} we decompose the measured total fluxes
into stellar and excess near-IR fluxes, an essential step before the
visibility data for each source can be modelled in $\S$~\ref{models}.
In $\S$~\ref{conclusions} we summarize our results, and consider the
ensemble of data in order to discuss the implications of our results for
the nature of the CS environment in HAEBE stars.
 
\section{EXPERIMENTAL PROCEDURE AND DATA REDUCTION} \label{exp}

	\subsection{Experimental Procedure}

The observations described in this paper were carried out at the IOTA,
a Michelson stellar interferometer located on Mount Hopkins, Arizona
(see \citealt{Traub:1998} for a recent review of the IOTA instrument).
Observations were made in the near-IR H ($\lambda_0 = 1.65 \mu m,
\Delta \lambda = 0.30 \mu m$) and K$'$ ($\lambda_0 = 2.16 \mu m,
\Delta \lambda = 0.32 \mu m$) bands; and using two IOTA baselines, of
lengths $B = 21$ m (North-South orientation) and 38 m (North-North
East orientation). For reference, the resolution corresponding to the
longest baseline, as measured by the full-width at half maximum (FWHM)
of the response to a point source, is $\lambda/2B = 4$ mas (1.8 AU)
and 6 mas (2.8 AU) at H and K$'$ respectively, where the linear
separations are given for a median distance of 460 pc to the stars in
our sample.

At the IOTA, starlight is collimated into $4.5\,$cm diameter beams 
by the telescope assemblies, of aperture diameter $D=45\,$cm, 
and are directed in vacuum towards the delay line
carriages. The delay lines serve to maintain the zero optical path
difference (OPD) condition between the two arms of the interferometer,
from the source to the beam combination point, by reflecting the beam
that requires a delay onto movable dihedral mirrors which track the
sidereal motion of the target star.  The path-equalized beams then
emerge from the vaccuum tank and propagate in air inside the beam
combination laboratory. In this experiment, a pair of dichroics
reflect the near-IR light towards an optical table containing the beam
combination optics and fringe detection system. The visible light in
the beams is transmitted toward the CCD-based wavefront tip-tilt 
compensation servo system.

Interference is produced in the pupil plane, by combining the
collimated and path-equalized beams at a beam splitter.  The two
outputs of the beam splitter are focused onto two separate pixels of a
NICMOS3 array, and a scan containing an intensity fringe packet is
recorded at each pixel by modulating the OPD by $\pm 60\,\mu m$ via an
extra reflection in one of the arms at a mirror mounted on a piezo
stack driven by a highly linear triangle waveform. The scan rate is
selectable in the range $1-10$ scans/sec.

Each scan contains 256 samples, and the integration time per data
point is in the range $0.18-2.0\,$ms, chosen as the optimum in order
to achieve maximum flux sensitivity while keeping the scan time short
in order to freeze the wavefront piston fluctuations induced by the
atmosphere during the fringe packet acquisition.  With this detection
system, the limiting magnitudes for real-time visual detection of a
fringe packet in a single scan are 7 at H-band (read-noise limited)
and 6.2 at K$'$-band (background limited), given here as the
magnitudes of the faintest sources observed under typical conditions
\citep{Millan:1999a,Millan:1999c}.

A typical observation consists of 500 scans obtained in 1-8 min,
followed by 10 scans on the sky which are used to measure the
background flux.  Target observations are interleaved with an
identical sequence obtained on an unresolved star, which serves to
calibrate the interferometer's instrumental response and the effect of
atmospheric seeing on the visibility amplitudes. The target and
calibrator sources are typically separated on the sky by a few degrees,
and observed a few minutes apart; these conditions insure that the
calibrator observations provide a good estimate of the instrument's
transfer function.

	\subsection{DATA REDUCTION AND CALIBRATION} \label{datared}

	\subsubsection{Fringe Fitting}

The data reduction process begins by estimating the fringe visibility,
i.e. the fringe amplitude relative to the background-subtracted mean
intensity in each interferogram. The raw response from the NICMOS3 pixels
is first corrected for their intrinsic non-linearity, for both star
and sky scans. A ``median sky scan'' is then computed as a
point-by-point median of the 10 individual sky scans, and subtracted
point-by-point from the star scans.  If we call the resulting scans
$I_A$ and $I_B$ for each side of the beam splitter, we then compute
{\it reduced scans} according to

\begin{equation}
I = \frac{I_A/\bar{I}_A - I_B/\bar{I}_B}{I_A/\bar{I}_A + I_B/\bar{I}_B}
\end{equation}

\noindent where $\bar{I}$ denotes the mean. The above operation: (1)
compensates for any existing photometric unbalance between pixels A
and B; (2) improves the SNR, since the beam splitter outputs $I_A$ and
$I_B$ are out of phase by $\pi$, while the noise is uncorrelated; and
(3) eliminates common-mode noise, most notably intensity fluctuations in
the signal due to scintillation and 
residual seeing-induced motion of the focused star images 
outside the NICMOS3 pixels.

The central 3 fringes in the reduced scans are then fitted in the time
domain to a point source response template computed as the Fourier
transform of the spectral bandpass used in the observation (H or
K$'$).  As a result of the fringe fitting procedure, we obtain the
following quantities for each reduced scan: mean, fringe amplitude,
position of the central fringe in the scan, and sampling rate. The
data are flagged as being a false fringe identification if an
individual fringe visibility is outside the range $0-100\%$, or if
either the central fringe position or sampling rate has a value which
is more than three times the standard deviation obtained from the
ensemble of the 500 reduced scans in the observation. Finally, an
entire observation is rejected if the standard deviation of the
central fringe position values exceeds $L/2$, where $L=120\,\mu$m is
the optical path corresponding to the scan length.

	\subsubsection{Visibility Estimation}

Propagation through the Earth's atmosphere and inside the instrument
degrades the interfering wavefronts differentially from their initial
perfectly plane shapes, and therefore reduces the fringe visibility
that is measured. The instrumental terms are constant or slowly
changing and can therefore be monitored and calibrated using
observations of a reference star known to be unresolved or of well
known angular diameter.  The atmospheric terms, however, change on
time scales much shorter than the time required to do an observation
and therefore cannot be calibrated using observations of a reference
star. In this section we address the problem of estimating the
visibility for an observation in a way that is unbiased by low-order
atmospheric effects.

\begin{figure}[htbp]
\begin{center}
\includegraphics[height=3in]{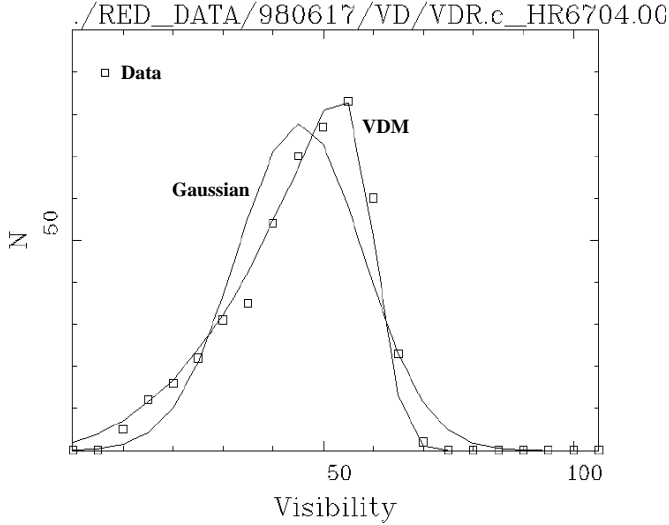}
\caption{Comparison of visibility data in an observation (500
scans) with a Gaussian distribution and the visibility distribution
model (VDM). In this case the visibility estimate given by the mean is
$42.8 \pm 0.6$\%, while the estimate given by the VDM is $V = 60.9 \pm
0.3$\%.  The other VDM parameters obtained are: $\Delta \phi_r = 0.15$
rad, $\sigma_{\Delta \phi_r} = 0.15$ rad and $\sigma_{V_{noise}} =
4$\%.\label{vdm}}
\end{center}
\end{figure}

A histogram of visibility values obtained in a typical observation is
shown in Figure~\ref{vdm}, along with a curve representing the
Gaussian distribution with the measured mean and variance. It can be
clearly seen that the data distribution is not well represented by the
Gaussian distribution, so that traditional estimators such as the
mean would not be appropriate. For this work, we have developed a new
technique which relies on our capability to record hundreds of
individual fringes in only a few minutes, so that the distribution of
visibilities is well determined and used in the estimation process.
The size of the IOTA telescopes is well matched to the area over
which an atmospherically disturbed wavefront is expected to remain
approximately flat, nominally $r_0 \sim 50\,cm$ in the
near-IR. Therefore, the atmospheric visibility degradation is likely
to be dominated by the residual differential wavefront tilts
uncompensated for by the tip-tilt servo. Consequently, we consider the
following representation for the measured visibilities

\be
V_{measured}^i = V_{true} \cdot V_{inst} \cdot
\left[ V_{tilt}^i + V_{noise}^i  \right];
\quad i=1, \ldots ,n
\ee

\noindent where $V_{true}$ is the true object visibility, 
$V_{inst}$ represents the effect of slowly changing instrumental
terms, such as wavefront curvature or polarization mismatch, 
$V_{tilt}^i$ is the visibility that results from interfering two
tilted wavefronts, $V_{noise}^i$ results from the fringe
fitting process in the presence of read and photon noise, and
$n=500$ is the number of individual visibility values.  Considering
linear phase gradients on each aperture of differential slopes $\Delta
\phi_{x,y}^i$ along orthogonal directions, it is straightforward to
show that the resulting tip-tilt term is given by

\be
V_{tilt}^i = \frac{J_1(2 \pi \Delta \phi_r^i)}{\pi \Delta \phi_r^i}; \quad
\mbox{with:} \quad
\Delta \phi_r^i = \sqrt{(\Delta \phi_x^2 + \Delta \phi_y^2)^i}
\ee 

A realization of the visibility distribution is generated by drawing
values, $n$ times, of $\Delta \phi_{r}^i$ and $V_{noise}^i$ from
Gaussian distributions of [mean,variance] equal to [$\Delta
\phi_r$~,~$\sigma_{\Delta \phi_r}^2$] and [0~,~$\sigma_{V_{noise}}^2$]
respectively. The physical motivation is that the mean differential
tilt ($\Delta \phi_r$) represents a fixed instrumental misalignment,
affected by seeing-induced fluctuations of variance $\sigma^2_{\Delta
\phi_r}$.

For a each observation, an estimate of the product $V=V_{true} \cdot
V_{inst}$ is obtained by finding the best match, in a $\chi^2$ sense,
between the data distribution and the above model, in a grid generated
for the complete range of the 4 parameters: $V_{true}$,
$\sigma_{V_{noise}}$, $\Delta \phi_r$ and $\sigma_{\Delta \phi_r}$.
In practice, due to the fact that in this model it is difficult to
distinguish the effects of the true visibility and mean tilt
parameters on the distribution, the value of the mean tilt for a given
night and for a given experimental set-up is obtained from the mean of
preliminary fits to all the calibrator observations, and fixed to this
value in subsequent fits of the target and calibrator observations.
Similarly, the noise parameter is constrained for each observation to
be the RMS in the visibilities obtained in the fringe fitting process.
An example of the best-fit model distribution found is also shown in
Figure~\ref{vdm}.

	\subsubsection{Calibration}

A spline interpolation between sequential $V$ estimates obtained in
observations of calibrator sources, for which $V_{true}$ is known,
yields a curve which reflects the time evolution of the instrument.
Interleaved target observations, taken under identical instrumental
conditions, are calibrated for the $V_{inst}$ term by dividing the
target estimate by the interpolated calibrator estimate:
$V_{true}^{target}=V_{true}^{cal} \cdot V^{target}/V^{cal}$.  The
spectral types and visual magnitudes of all the calibrator stars used
in this paper (with the possible exception of HR8881, see
section~\ref{sample}) have angular diameters less than $1-2$ mas,
therefore, we use $V_{true}^{cal}=1.0$.

	\subsubsection{Errors}

\begin{figure}[htbp]
\begin{center}
\includegraphics[height=3in]{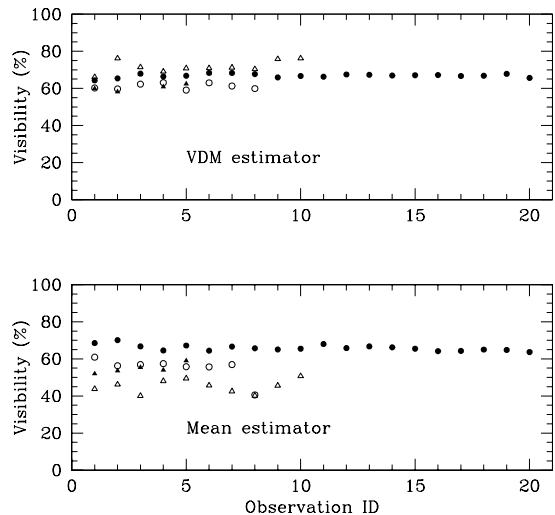}
\caption{Calibrated visibilities ($V_{true}$) obtained using the
visibility distribution model (top panel) and the mean of the measured
distribution (bottom panel).  Each symbol represents data taken on the
same source (AB Aur) on different nights covering a large range of
seeing conditions, and using identical detection parameters. For
reference, the longest set of data (filled bullets) spans 5.2
hours.\label{abaurvdm}}
\end{center}
\end{figure}

The internal uncertainty from fitting individual fringes is generally
small ($\lesssim$1\%), as is the formal uncertainty in the visibility
for an observation obtained from the fit to the visibility
distribution model ($\lesssim 2-3$\%). However, the RMS of independent
visibility measurements on the same object and at similar spatial
frequencies is larger, typically $\lesssim$5\% p-p (peak-to-peak)
within a single night of observation, although larger, $\leq$15\% p-p
in H-band data and $\leq$20\% p-p in K$'$-band data, systematic
differences occasionally exist when comparing data from different
nights, both of which indicate the presence of low-frequency
calibration errors. A likely explanation for this excess error is that
our expectation that $r_0(H,K') \sim D$ does not strictly hold, and
that we see the effects of un-modelled wavefront curvature errors
induced by the turbulent atmosphere.  Therefore, the total uncertainty
on an individual visibility measurement due to all sources is best
represented by the RMS of independent measurements separated by hours
of days.  Using this measure, we find that the use of the visibility
distribution model yields more consistent results, reducing the
scatter of independent measurements by factors of $2-3$ compared to
the mean estimator.  This improvement is illustrated in
Figure~\ref{abaurvdm}, which compares both methods on calibrated
visibilities obtained on the same target on 4 nights of very different
seeing conditions.  We note that the night-to-night variations
observed in this example are not believed to be due to the object
itself (AB~Aur) , but are instrumental.  Variations in the object
visibility are expected if (1) the source is asymmetric on the sky, or
(2) it is composed of more than one component with changing relative
fluxes. However, during one of the nights this source has been
monitored for 5.2 hours (solid bullets in the plot), showing no sign
of variation and implying a high degree of circular
symmetry. Moreover, according to \citet*{Vosh:1996}, AB Aur is an
irregular photometric variable of very small amplitude, $\Delta m =
\pm 0.02$. In a representation of the source consisting of a central
star surrounded by near-IR emitting material such a small change in
the brightness of the central star would result in a change of the
observed visibility $< 1$\%, and therefore this effect is not likely
to be the cause of the observed night-to-night variations.

\vspace{-0.1in}

\clearpage

\section{THE SAMPLE, LOG OF OBSERVATIONS, AND VISIBILITY AMPLITUDES} 
\label{sample}

To date, the most complete and comprehensive compilation of candidate
HAEBE stars is the catalog of 287 objects by \citet*{The:1994}. Our
sample was selected from their Table 1, which contains 108 Ae and Be
stars recognized as true members or good candidates of the HAEBE
group, using as the selection criteria that the stars be brighter than
the limiting magnitudes of IOTA's tip-tilt servo system ($V=11-12$)
and near-IR fringe detector (H=7, K$'$=6.2).  Furthermore, in order to
have as homogenous a sample as possible, and in order to test the
accretion disk hypothesis, we have primarily selected our sources to
belong to Group~I in the classification of HSVK. The only exceptions
are $\omega$~Ori and MWC~166, which are Group~III stars in that
classification.

The final list of targets observed is presented in
Table~\ref{targetstable}.  Columns~$1-3$ contain the star names and
coordinates; columns~$4-10$ show the VRIJHK photometry compilation,
mostly from HSVK; columns~$11-15$ contain stellar parameters,
distance, visual extinction, spectral type and effective temperature,
mostly from the Hipparcos data study of \citet{Ancker:1998}; and the
last column comments on whether the star posseses a visual or
spectroscopic companion and/or IR excess. Table~\ref{obslog} shows a
log of the observations, including the calendar date, spectral band,
calibrator used, the approximate hour angle (HA) range covered during
the observation, a qualitative measure of the seeing at the time of
the observations (1=very good, 4=bad) based on real-time visual
inspection of the amount of atmospherically induced fringe piston and
amplitude noise, and the baseline used (identified by its approximate
length).

	\subsection{Special Cases and Visibility Correction Factors}

	\subsubsection{Stars with Close Visual Companions} \label{ao}

Using the technique of adaptive optics in the near-IR,
\citet{Corporon:1998} surveyed 68 HAEBE stars and identified 30 binary
systems with separations greater than 0.1 arcsec. We consider here the
effect of those ``wide'' companions on our data set.  Indeed, if a
star has a companion at an angular offset such that it is in the
detector's field of view ($\sim 3\arcsec \times 3\arcsec$ for each
pixel of the NICMOS3 camera), but whose interferogram has a zero OPD
point that does not fall near that from the primary source, in the
resulting scan a correction needs to be made to the fringe visibility
measured on each interferogram that accounts for the incoherent flux
added by the other star (exactly analogous to the incoherent flux
added by thermal background, which needs to be subtracted before the
fringe amplitude is measured). If we call $V_{true}^{j}$ the
calibrated visibility corresponding to the bright ($j=A$) or faint
($j=B$) fringe obtained by the procedure described in
Section~\ref{datared}, then the corrected visibility is given by

\be
V_{true}^{'j} = V_{true}^{j}/r_j \,;\quad \mbox{with:} \quad
r_{A,B} = \frac{1}{(1 + 10^{\mp \Delta m/2.5} )} 
\ee

\noindent where $\Delta m = m_B - m_A$ is the magnitude difference
between the binary components in the relevant spectral band.

Our sample overlaps almost perfectly with that of
\citet{Corporon:1998}, and therefore we use his measured binary
separations and flux ratios to apply the above correction in the
appropriate cases.  Table~\ref{adaptive} summarizes the angular
separations ($\theta$), magnitude differences and the spectral types
assigned to each component, for the subset of our sample for which the
adaptive optics companion is approximately within the field of view of
our NICMOS3 camera. Note that in our data set, the interferogram
corresponding to the faint companion is never detected, either because
it is too faint or because it occurs at a value of the OPD which is
outside of the scan length. However, the correction by the $r_A$
factors, last two columns in Table~\ref{adaptive}, still needs to be
applied to the fringe visibility corresponding to the bright
component. The only two stars in our sample which were not observed by
Corporon are AB~Aur and V594~Cas, and we will assume here that they
are single stars.

	\subsubsection{MWC 1080 and HR8881} \label{mwc1080}

The data for MWC~1080 shows that it is more resolved at H than at
K$'$, by an amount significantly greater than expected from the higher
resolving power of the shorter H-band wavelength.  In searching for
explanations for this unusual circumstance, it was discovered that
according to \citet{Hammersley:1994}, the calibrator used for MWC~1080
(HR~8881), is misclassified in the Bright Star Catalogue\footnote{The
Bright Star Catalogue, Fourth Revised Edition, Yale University
Observatory, 1982}, and is really a M3 giant or K supergiant, and not
a K1 dwarf.

Assuming that HR~8881 is a normal giant, we may estimate its angular
size and account for it in the calibration of the MWC~1080 data. Based
on a compilation of angular diameters measured in the near-IR by the
techniques of lunar occultations and long baseline inteferometry,
\citet{vanBelle:1999} has derived empirical relations between angular
diameters and $(V-K)$ and $(B-K)$ colors for main sequence, giants and
supergiants stars.  Using the photometry and $A_V=0.22$ estimate of
\citet{Hammersley:1994}, we obtain $(V-K)=4.05 \pm 0.02$ and
$(B-K)=5.33 \pm 0.02$. The empirical relations using those two colors
predict uniform disk (UD) diameters $\theta=2.12 \pm 0.28$ mas and
$\theta=2.08 \pm 0.37$ mas respectively. The weighted mean of the two
predictions is $\theta = 2.10 \pm 0.22$ mas, which we adopt as our
estimate of the angular size of HR~8881.

The visibility data for MWC~1080 has therefore been calibrated using
the visibility that is expected for the estimated UD diameter
$V_{cal}^{UD} = 2 \cdot J_1(\pi \theta s)/(\pi \theta s)$. For
reference, for the baseline lengths corresponding to those
obervations, this calibration factor is on average 0.95 for the H band
observations and 0.97 for the K$'$ band observations.

The unusual circumstance therefore remains that MWC~1080 is resolved
by about 20\% in H band observations and appears essentially
unresolved in K$'$ observations. This difference is difficult to
explain with any simple model of the source, given that the near-IR
excess is greater at K than at H, contributing 79\% and 91\% of the
total flux at H and K respectively (see Section~\ref{fluxes}).  One
possible explanation would be that the calibrator, HR~8881, is not a
UD, and that the interferometer resolves it at K$'$ by a greater
amount than is predicted by its photospheric extension. There are
however no specific measurements in the literature to confirm or rule
out this hypothesis, and a complete resolution to this question will
have to await new observations at the IOTA of both MWC~1080 and
HR~8881.  If the result is confirmed, the apparent decrease in size
with increasing wavelength could be explained if MWC~1080 is an
under-resolved core-halo object. In this model, the small core
contains the central star as well as the source of near-IR excess and
dominates the emission at K$'$ wavelengths, while at the shorter H
wavelengths scattering reveals the larger angular extent of the cold
halo component.

	\subsection{Visibility Data}

\begin{figure}[htbp]
\begin{center}
\includegraphics[height=7in]{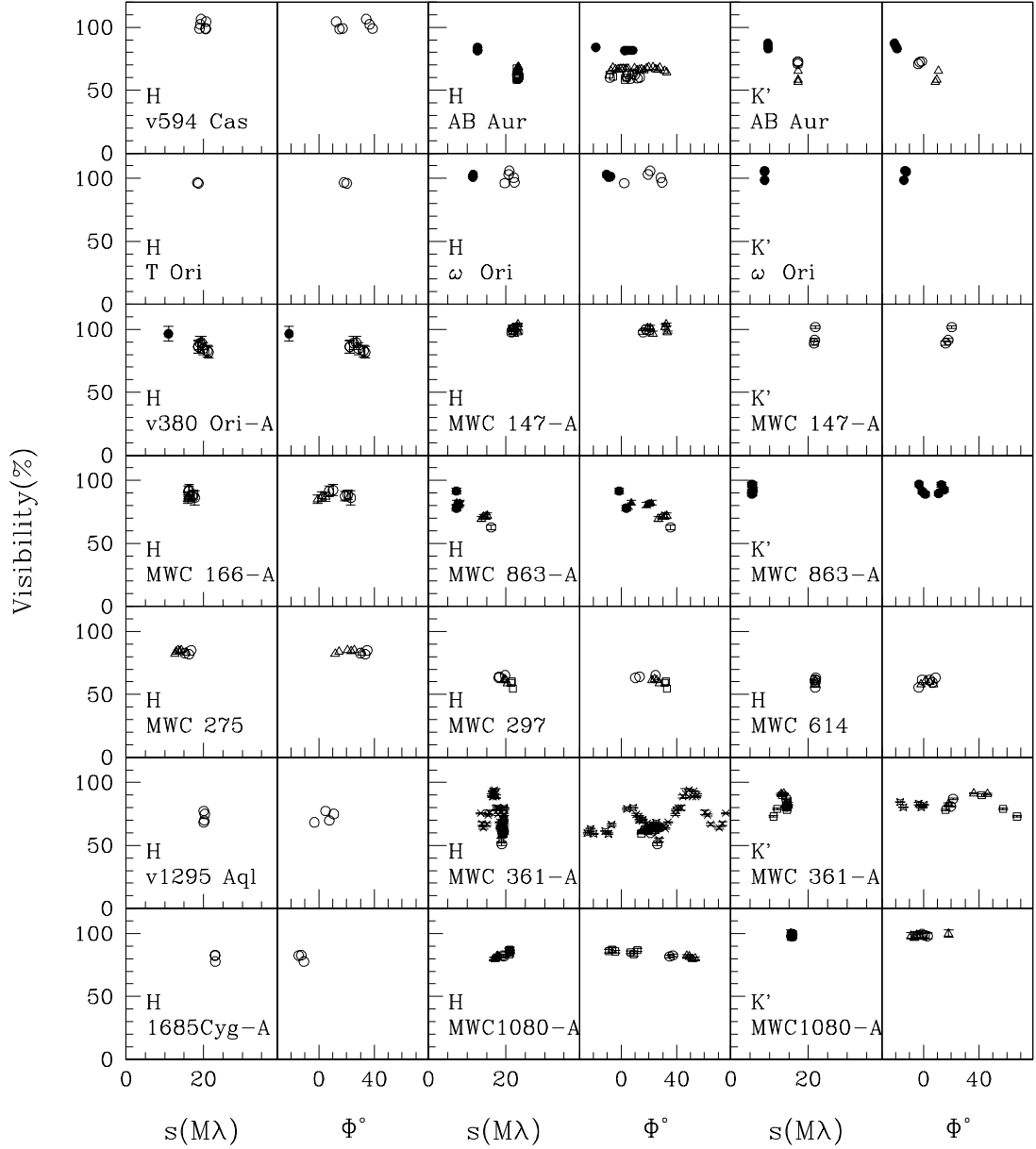}
\caption{H and K$'$-band calibrated visibility data. For each
source, the data are plotted as a function of the projected baseline
length ($s$, left panel) and position angle ($\Phi$, right panel).
The different symbols correspond to observations on different nights,
and retain their meaning between the right and left panels for a given
source. Observations on the 21~m baseline are represented with filled
symbols, and observations at the 38~m baseline with open symbols. All
subsequent plots involving this visibility data retain the same
symbols.\label{visdata}}
\end{center}
\end{figure}

The final set of calibrated H and K$'$-band fringe visibility data is
shown in Figure~\ref{visdata}.  The data are presented in two panels,
one for each independent variable specifying the projection of the
baseline vector in a plane ($u,v$) normal to the direction toward the
source. The left panels show the data as a function of the modulus of
the projected baseline: $s = \sqrt{u^2 + v^2}$, expressed in millions
of wavelengths at the center of the observation band.  The left panels
show the data as a function of the projected baseline position angle:
$\Phi = \arctan \left(u/v\right)$, measured from North toward East.
The size of the data symbols equals our typical uncertainty of 5\% in
the calibrated visibilities. In the case of the stars known to be
adaptive optics binaries, only the bright component (A) detected
interferometrically appears in the figure, and explicit error bars are
shown reflecting the error in correction by the flux factor described
in Section~\ref{ao}.  The error bars in the data for MWC~1080 also
include the contribution due to the uncertainty in the estimated size
of HR~8881.

The detailed interpretation of this data set is the subject of the
next two sections.  However, based on inspection, we may immediately
make the following observations: (1) of the 15 stars observed, 11
(73\% of the sample) are resolved by the interferometer; (2) one star,
MWC~361-A, shows the clear sinusoid signature of a binary system; (3)
none of the other stars shows significant variation of visibility with
baseline position angle (the range covered by the observations varies
from $\Delta \Phi = 0$ to $65\arcdeg$); and (4) of the two Group III
sources, $\omega$~Ori appears unresolved, and MWC~166 appears only
slightly resolved, both consistent with that classification.

\addtocounter{table}{+1}
\begin{figure}[htbp]
\begin{center}
\includegraphics[angle=180,scale=0.85]{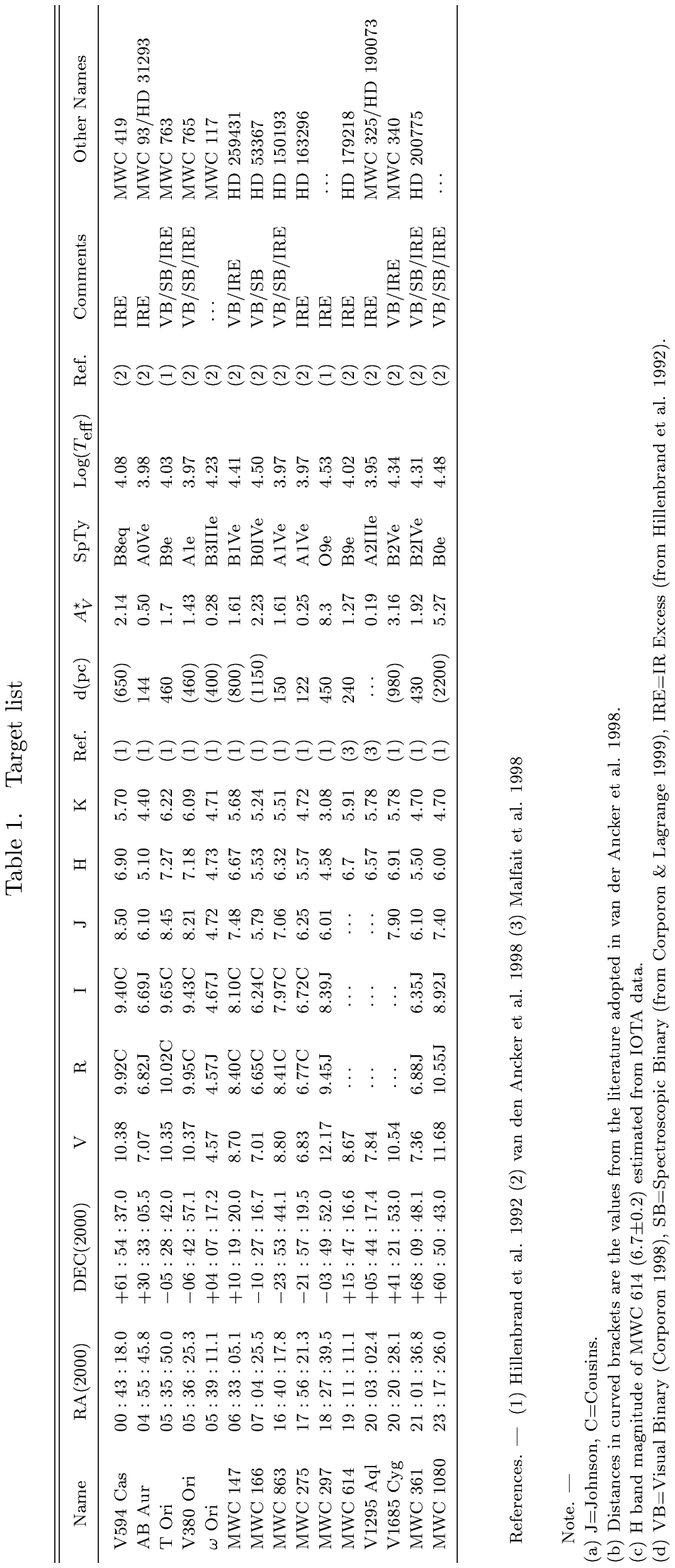}
\label{targetstable}
\end{center}
\end{figure}
\nocite{Hillenbrand:1992}
\nocite{Malfait:1998}
\nocite{Ancker:1998}



\begin{deluxetable}{llclccc}
\tablecaption{Log of Herbig Ae/Be Stars Observations \label{obslog}}
\tablewidth{0pt}
\tablehead{
\colhead{Date}  & \colhead{Target} & \colhead{Band} & 
\colhead{Calibrator} &
\colhead{HA} & \colhead{Seeing\tablenotemark{a}} & \colhead{Baseline} \\
& & & \colhead{(HR)} & & & (m)
} 
\startdata
1997 Oct 13 & AB Aur   & H   & 1343,2219 & $[-1.2 , +0.6]$         & 2 & 21  \\
1997 Nov 15 & AB Aur   & H   & 1626      & $[-1.0 , +3.0]$   & 1 & 38  \\
1998 Jan 19 & AB Aur   & K$'$ & 1626     & $[+0.9,+1.1]$     & 2 & 38 \\	   
1998 Mar 02 & AB Aur   & K$'$  & 1626         & $[+1.9 , +2.1]$   & 1 & 21    \\
1998 Mar 03 & $\omega$ Ori    & H   & 2019      & $[+0.3 , +0.7]$   & 2 & 21  \\
         &          & K$'$  & 2019         & $[+0.9 , +1.1]$   & 2 & 21    \\
1998 Mar 07 & AB Aur   & H   & 1626      & $[+1.6 , +1.9]$   & 2 & 38  \\
         &          & K$'$  & 1626         & $[+2.2 , +2.5]$   & 2 & 38    \\
1998 Jun 12 & MWC 297  & H   & 7149      & $[+0.9 , +1.2]$   & 3 & 38  \\
         & MWC 361  & H   & 7967      & $[-0.4 , +0.2]$   & 3 & 38     \\
1998 Jun 13 & MWC 863  & H   & 6153      & $[-0.8 , -0.5]$   & 2 & 38 \\
	 & MWC 275  & H   & 6704      & $[-0.9 , -0.1]$   & 2 & 38    \\
         & MWC 361  & H   & 7967      & $[+0.4 , +0.7]$   & 1 & 38    \\
1998 Jun 14 & MWC 297  & H   & 7149      & $[-0.8 , -0.0]$   & 2 & 38  \\
1998 Jun 15 & MWC 361  & H   & 7967      & $[-0.5 , -0.0]$   & 1 & 38  \\
1998 Jun 17 & MWC 863  & H   & 6153      & $[-0.1 , -0.5]$   & 1 & 38  \\
         & MWC 275  & H   & 6704      & $[-0.3 , +1.4]$   & 1 & 38    \\
         & MWC 361  & H   & 7967      & $[-0.5 , +0.7]$   & 1 & 38      \\
1998 Jun 18 & MWC 297  & H   & 7066      & $[-1.9 , -1.6]$   & 1 & 38  \\
         & MWC 361  & H   & 7967      & $[-0.8 , -0.2]$   & 1 & 38    \\
         &          & K$'$  & 7967         & $[+0.0 , +0.2]$   & 1 & 38    \\
         & MWC 1080 & H   & 8881      & $[-1.8 , -1.4]$   & 1 & 38    \\
1998 Jun 19 & MWC 361  & H   & 7967      & $[-2.8 , +1.0]$   & 1 & 38  \\
         &          & K$'$  & 7967         & $[-2.1 , +0.4]$   & 1 & 38    \\
1998 Jun 20 & MWC 361  & H   & 7967      & $[-4.9 , +0.4]$   & 1 & 38  \\
         &          & K$'$  & 7967         & $[-4.3 , +0.6]$   & 1 & 38    \\
         & MWC 1080 & H   & 8881      & $[-3.5 , -2.9]$   & 1 & 38    \\
1998 Sep 28 & MWC 361  & H   & 7967      & $[-0.7 , +4.4]$   & 1 & 38  \\
         &          & K$'$  & 7967         & $[+2.1 , +3.7]$   & 1 & 38    \\
1998 Sep 29 & V1295 Aql & H  & 7669      & $[+1.0 , +2.4]$   & 2 & 38  \\
1998 Sep 30 & MWC 614  & H   & 7449      & $[+1.2 , +2.4]$   & 1 & 38  \\
1998 Oct 02 & MWC 614  & H   & 7449      & $[+1.3 , +2.2]$   & 1 & 38  \\
         & V1685 Cyg & H  & 7800      & $[+2.4 , +3.7]$   & 2 & 38    \\
1998 Oct 31 & MWC 1080 & H   & 8881      & $[+0.9 , +2.9]$   & 1 & 38  \\
         &          & K$'$  & 8881         & $[+1.8 , +3.8]$   & 1 & 38  \\
         & T Ori    & H   & 1986      & $[-0.0 , +0.4]$   & 1 & 38  \\
1998 Nov 02 & AB Aur   & H   & 1626      & $[-2.3 , +2.9]$   & 1 & 38  \\
1998 Nov 03 & V594 Cas & H   & 237       & $[-1.9 , +0.8]$   & 1 & 38  \\
         & V380 Ori & H   & 1697      & $[-1.8 , -0.4]$   & 1 & 38    \\
         & MWC 147  & H   & 2426      & $[-0.2 , +0.4]$   & 1 & 38    \\
         &          & K$'$  & 2426         & $[+0.8 , +1.3]$   & 1 & 38    \\
1998 Nov 07 & $\omega$ Ori    & H   & 2019      & $[-1.4 , +2.0]$   & 3 & 38 \\    
1998 Nov 10 & MWC 1080 & K$'$  & 8881         & $[+0.0 , +3.0]$   & 2 & 38    \\
         & MWC 147  & H   & 2426      & $[-2.5 , -0.0]$   & 2 & 38    \\
         & MWC 166  & H   & 2723      & $[+0.1 , +1.5]$   & 3 & 38    \\
1999 Mar 01 & V380 Ori & H   & 1986      & 1.9	       	  & 2 & 21  \\
	 & MWC 863  & H   & 6153      & $[-0.8 , -0.4]$	  & 2 & 21    \\
1999 Mar 02 & MWC 863  & H   & 2723      & $[-2.3 , -0.9]$   & 2 & 21    \\
         &          & K$'$  & 2723         & "	          & 2 & 21    \\
1999 Mar 05 & MWC 166  & H   & 2723      & $[+1.7 , +2.2]$   & 2 & 38  \\
\enddata

\tablenotetext{a}{The qualitative measure of seeing goes from ``good'' = 1 
	to ``bad'' = 3}
\end{deluxetable}



\begin{deluxetable}{llllllllcccc}
\tablecaption{HAEBE stars with adaptive optics companions \label{adaptive}}
\tabletypesize{\small}
\tablecolumns{12}
\tablewidth{0pt}
\tablehead{ 
\colhead{Name} & \colhead{$\theta \arcsec$} & \colhead{$\Delta V$} & 
\colhead{$\Delta R$} & \colhead{$\Delta I$} & \colhead{$\Delta J$} & 
\colhead{$\Delta H$} &
\colhead{$\Delta K$} & \colhead{SpTy} & \colhead{SpTy} &
$r_A(H)$ & $r_A(K')$ \\
& & & & & & & & A & B & &
}
\startdata
V380 Ori  	& 0.14	& \nd	& 0.10	& \nd  & 0.11 & 0.94 & 1.36 & B7 & G0 &	
	0.704 $\pm$ 0.04\phn & 0.778 $\pm$ 0.03\phn \\
MWC 147		& 3.1	& 7.22	& 6.82  & 6.02 & 4.81 & 5.05 & 5.67 & B2 & K5 & 
	0.990 $\pm$ 0.002 & 0.995 $\pm$ 0.001 \\
MWC 166		& 0.64	& 1.43  & 1.61  & 1.7  & 1.70 & 1.40 & 1.77 & B0 & B0 & 
	0.784 $\pm$ 0.03\phn & 0.836 $\pm$ 0.02\phn \\
MWC 863		& 1.07	& 3.57  & \nd   & 2.57 & 2.07 & 2.24 & 2.72 & A2 & K4 & 
	0.887 $\pm$ 0.02\phn & 0.924 $\pm$ 0.01\phn \\
V1685 Cyg	& 0.72	& \nd   & \nd   & \nd  & \nd  & \nd  & 5.50 & B2 & \nodata & \nodata & 
			0.994 $\pm$ 0.001 \\
MWC 361		& 2.55	& 5.40  & 5.04  & 4.52 & \nd  & 4.12 & 4.64 & B2 & G5 & 
	0.978 $\pm$ 0.004 & 0.986 $\pm$ 0.002 \\
MWC 1080	& 0.77	& 2.34  & 2.44  & 2.61 & \nd  & 2.84 & 2.28 & B2 & B2 & 
	0.932 $\pm$ 0.01\phn & 0.891 $\pm$ 0.02\phn \\
\enddata
\tablerefs{Angular separation, photometric and spectral type data from 
\citet{Corporon:1998}.}
\tablecomments{
(1) Completeness limits: $ 0.1 \arcsec \lesssim \theta \lesssim 7 \arcsec$; 
	$\Delta m < 6-7$; 
(2) Uncertainties: $\sigma_{\theta} = 0.05 \arcsec$;
$\sigma_{\Delta m} \lesssim 0.2$. 
}
\end{deluxetable}

\clearpage

\section{EXCESS NEAR-INFRARED FLUX} \label{fluxes}

Most of the HAEBE stars in this paper are known to have a strong
near-IR excess, and are resolved by the interferometer.  Therefore, in
order to interpret the visibility amplitude data we consider a two
component model consisting of the central star, plus a component which
is the source of the near-IR excess. However, before this model can be
used to interpret the interferometer data, the relative contributions
to the total H and K fluxes from these two components must be
quantified.  The notation used hereafter is as follows: ``$T$'',
``$\star$'' and ``$E$'' superscripts denote total, stellar and excess
fluxes respectively. Primed fluxes are apparent, while un-primed
fluxes are intrinsic (i.e. corrected for extinction).

	\subsection{Stellar Flux}

In order to estimate the stellar near-IR fluxes, we use the photometry
data, visual stellar extinctions and stellar effective temperatures
from the literature tabulated in Table~\ref{targetstable}. We assume
that the short wavelength (VRI) flux arises entirely from the central star,
and approximate the stellar spectrum as a blackbody at the adopted
effective temperature. The solid angle subtended by the stellar disk
of radius $R_{\star}$ at distance $d$, $\Omega_{\star} = \pi
R_{\star}^2/d^2$, is the parameter adjusted in order to provide a best
fit to the de-reddened V, R and I fluxes, according to:
$F_{VRI}^{\star}=B_{VRI}(T_{eff}) \cdot \Omega_{\star}$, where
$B_{\nu}(T)$ denotes the Planck function. The literature flux data has
been de-reddened using the extinction law ($A_{\lambda}/A_V$) of
\citet*{Steenman:1991} and a ratio of total to selective extinction
$R=A_V/E_{B-V}=3.10$.  Once the solid angle factor is thus determined,
the stellar flux may be calculated at any frequency as
$F_{\nu}^{\star} = B_{\nu}(T_{eff}) \cdot \Omega_{\star}$, and used to
quantify the flux in excess of the photospheric value.

The literature flux data is quoted as having uncertainties of 0.01$-$0.02 
mag in the VRI bands. In addition, the
spectral classification is uncertain by 1 sub-class typically, which
results in about 0.2 mag uncertainty in the derived values of the
visual extinction and about 11\% uncertainty in the effective
temperatures. The resulting relative uncertainties in the predicted near-IR
stellar fluxes are in the range $17\% - 50\%$ (median 30\%). 

In the case of those sources which were identified as adaptive optics
binaries by Corporon, we calculate the SED of each component by
combining the magnitude differences of Table~\ref{adaptive} with the
total photometry of Table~\ref{targetstable}.  We determine effective
temperatures for each component using the spectral types assigned by
Corporon and the temperature scale of \citet*{Cohen:1979}. These
effective temperatures are used to calculate the blackbody model which
represents the photospheric contribution of each star to the total
flux, as described above for the single stars.

	\subsection{Excess Flux}

Using this model to represent the stellar photosphere, we may deduce
the apparent excess fluxes due to CS emission: $F'^E_{JHK} =
F'^T_{JHK} - F^{\star}_{JHK} \cdot 10^{-A^{\star}_{JHK}/2.5}$, where
the apparent total fluxes and visual stellar extinctions are the
observed quantities from the literature.  Estimating the intrinsic
(i.e. de-reddened) excess fluxes is also of interest, since they
directly relate to the physical properties of the source. This
however, requires knowledge of the amount of attenuation suffered by
the excess radiation between the source and the observer, while the
only information available is the total extinction in the line of
sight toward the star. Therefore, we adopt here the simplifying
assumptions that (1) all of the stellar extinction is due to cold
foreground material, and (2) it is the same for all lines of sight
that intercept the extended component, so that
$F^{E}_{JHK}=10^{+A^{\star}_{JHK}/2.5} \cdot F'^E_{JHK}$.

\begin{figure}[htbp]
\begin{center}
\includegraphics[height=8.8in]{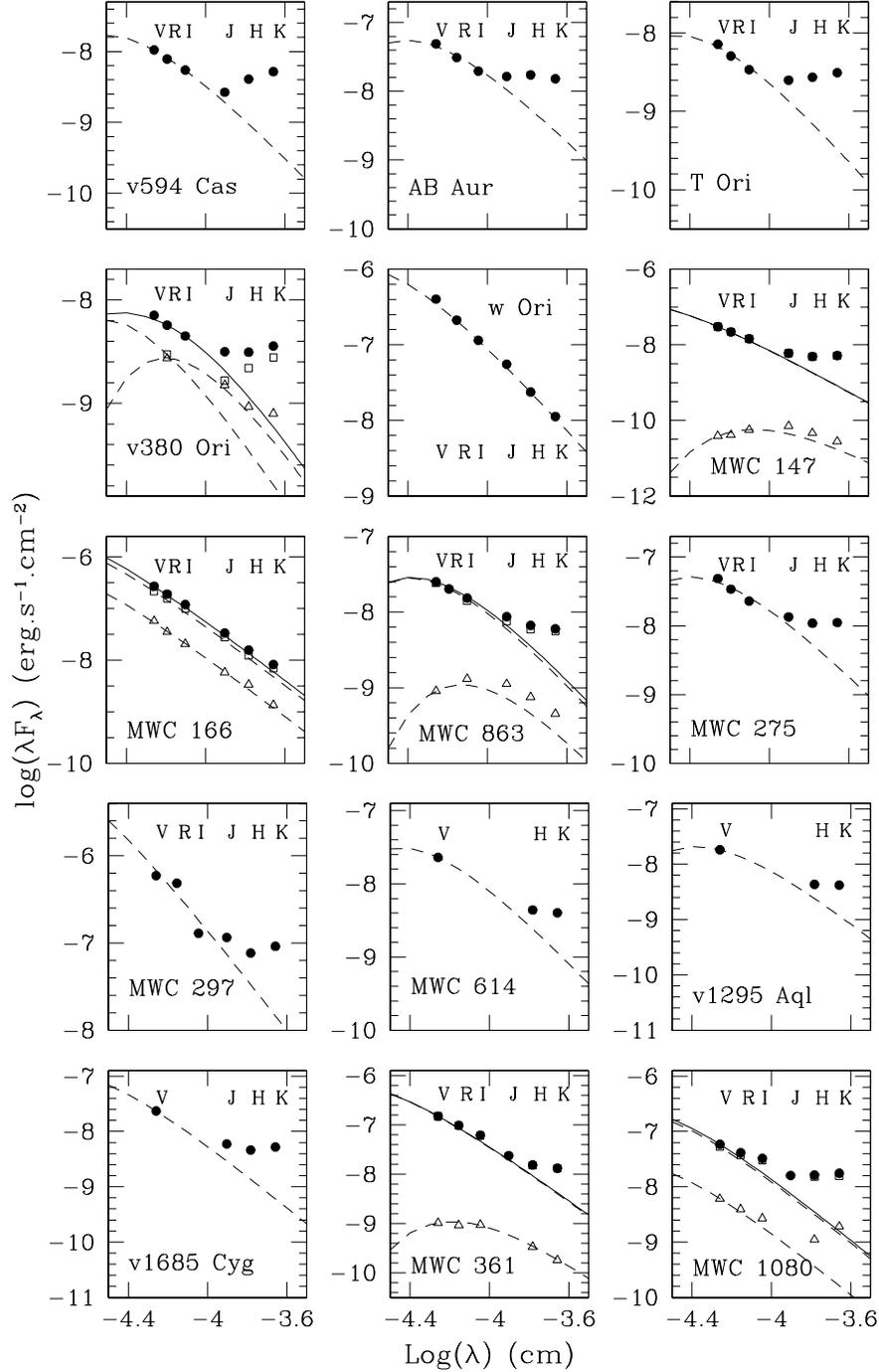}
\vspace{-1.3in}
\caption{Total SED data and photospheric models.  Each panel shows
the de-reddened total SED data of a target (circles) and the stellar
blackbody model (dashed lines) corresponding to the underlying star.
In the case of the adaptive optics binaries (Corporon 1998), the SED
data for each component is also shown (open squares and triangles) and
the solid line represents the sum of the two model photospheres.
\label{seds}}
\end{center}
\end{figure}

Figure~\ref{seds} shows the de-reddened SED data and best fit stellar
blackbody models for all the target stars. Note that in most cases the
individual components in the adaptive optics binaries also have
significant IR excess.  

The results for the derived H and K stellar fluxes are summarized in
Table~\ref{starfluxes}.  In addition to the quantities already
defined, the fourth column lists the stellar angular diameters that
are implied by the solid angle calculated, $\theta^{\star} =
2\sqrt{\Omega/\pi}$, and it can be seen that the values obtained are
all small compared to the resolution of the interferometer. Therefore,
in our models of the brightness distribution, the central stars will
be treated as point sources.

Table~\ref{excesstable} summarizes the results obtained for the
excess fluxes.  The errors are dominated by the uncertainty in the
determination on the underlying stellar spectrum, as described in the
previous section, and include an uncertainty of $0.01-0.1$~mag
from the total near-IR magnitudes in the literature, as well as a small
contribution due to the uncertainty in the extinction at near-IR
wavelengths derived from the visual extinctions. The last column in
Table~\ref{excesstable} contains the color temperature ($T_c$) of the
excess emission, calculated from the intrinsic H and K excesses such that
$F^{E}_{H}/F^{E}_{K} = B_H(T_c)/B_K(T_c)$.  

\begin{deluxetable}{llclllllll}
\tabletypesize{\footnotesize}
\tablecolumns{10}
\tablewidth{0pt}
\tablecaption{Near-IR stellar fluxes \label{starfluxes}}
\tablehead{\colhead{Name} & \colhead{$\mbox{log}(T_{\mbox{eff}})$} & 
	\colhead{$\Omega_{\star}$} &
	\colhead{$\theta_{\star}$} &
	\colhead{$F^{\star}_H$} & \colhead{$F^{\star}_K$} & 
	\colhead{$A_H$} & \colhead{$A_K$} & 
	\colhead{$F'^{\star}_H$} & \colhead{$F'^{\star}_K$} \\
 & & \colhead{($10^{-20}$ sr)} & (mas) & \colhead{(Jy)} & \colhead{(Jy)} & 
 & & \colhead{(Jy)} & \colhead{(Jy)}
}
\startdata
V594 Cas	& 4.08 & \phn 6.3 & 0.06 & \phn 0.5 $\pm$ 0.1 & \phn 0.3 $\pm$ 0.1 &
		 0.3 & 0.2 & \phn 0.4 $\pm$ 0.1 & \phn 0.3 $\pm$ 0.1 \\
AB Aur		& 3.98 & 50.3 & 0.2 \phn & \phn 3.0 $\pm$ 1.1 & \phn 1.9 $\pm$ 0.7 &
		 0.07 & 0.04 & \phn 2.8 $\pm$ 1.0 & \phn 1.8 $\pm$ 0.7 \\
T Ori	 	& 4.03 & \phn 5.4 & 0.05 & \phn 0.4 $\pm$ 0.1 & \phn 0.2 $\pm$ 0.1 & 
		 0.2 & 0.1 & \phn 0.3 $\pm$ 0.1 & \phn 0.2 $\pm$ 0.1 \\
V380 Ori-A 	& 4.09 & \phn 2.2 & 0.03 & \phn 0.2 $\pm$ 0.05 & \phn 0.1 $\pm$ 0.03 &
		 0.2 & 0.1 & \phn 0.1 $\pm$ 0.04 & \phn 0.1 $\pm$ 0.03 \\
$\omega$ Ori 	& 4.23 & 98.0 & 0.2 \phn & 12.9 $\pm$ 3.8 & \phn 7.8 $\pm$ 2.4 & 
		 0.04 & 0.02 & 12.4 $\pm$ 3.7 & \phn 7.6 $\pm$ 2.3 \\
MWC 147-A	& 4.31 & \phn 6.0 & 0.06 & \phn 1.0 $\pm$ 0.2 & \phn 0.6 $\pm$ 0.1 &
		 0.2 & 0.1 & \phn 0.8 $\pm$ 0.2 & \phn 0.5 $\pm$ 0.1 \\
MWC 166-A	& 4.49 & 22.0 & 0.05 & \phn 6.0 $\pm$ 1.3 & \phn 3.5 $\pm$ 0.8 &
		 0.3 & 0.2 & \phn 4.2 $\pm$ 0.9 & \phn 2.9 $\pm$ 0.6 \\
MWC 863-A	& 3.96 & 30.9 & 0.1 \phn & \phn 1.7 $\pm$ 0.6 & \phn 1.1 $\pm$ 0.4 & 
		 0.2 & 0.1 & \phn 1.3 $\pm$ 0.5 & \phn 1.0 $\pm$ 0.3 \\
MWC 275		& 3.97 & 51.2 & 0.2 \phn & \phn 2.9 $\pm$ 1.1 & \phn 1.9 $\pm$ 0.8 &
		 0.03 & 0.02 & \phn 2.8 $\pm$ 1.1 & \phn 1.8 $\pm$ 0.7 \\
MWC 297		& 4.53 & 60.7 & 0.2 \phn & 18.3 $\pm$ 3.4 & 10.6 $\pm$ 2.0 & 
		 1.2 & 0.7 & \phn 6.3 $\pm$ 1.2 & \phn 5.8 $\pm$ 1.1 \\
MWC 614		& 4.02 & 19.6 & 0.1 \phn & \phn 1.3 $\pm$ 0.4 & \phn 0.8 $\pm$ 0.3 & 
		 0.2 & 0.1 & \phn 1.1 $\pm$ 0.4 & \phn 0.8 $\pm$ 0.2 \\
V1295 Aql	& 3.95 & 24.9 & 0.1 \phn & \phn 1.3 $\pm$ 0.5 & \phn 0.9 $\pm$ 0.3 &
		 0.03 & 0.01 & \phn 1.3 $\pm$ 0.4 & \phn 0.8 $\pm$ 0.3 \\
V1685 Cyg	& 4.34 & \phn 4.1 & 0.05 \phn & \phn 0.7 $\pm$ 0.2 & \phn 0.4 $\pm$ 0.1 &
		 0.4 & 0.2 & \phn 0.5 $\pm$ 0.1 & \phn 0.4 $\pm$ 0.1 \\
MWC 361-A	& 4.31 & 29.5 & 0.1 \phn & \phn 4.9 $\pm$ 1.2 & \phn 2.9 $\pm$ 0.7 &
		 0.3 & 0.1 & \phn 3.6 $\pm$ 0.9 & \phn 2.5 $\pm$ 0.6 \\
MWC 1080-A	& 4.31 & 10.4 & 0.07 \phn & \phn 1.7 $\pm$ 0.4 & \phn 1.0 $\pm$ 0.2 &
		 0.7 & 0.4 & \phn 0.8 $\pm$ 0.2 & \phn 0.7 $\pm$ 0.2    
\enddata
\end{deluxetable}




\begin{deluxetable}{llllllllll}
\tabletypesize{\footnotesize}
\rotate 
\tablecolumns{10}
\tablewidth{0pt}
\tablecaption{Near-IR excess fluxes \label{excesstable}}
\tablehead{ \colhead{Name} & \colhead{$F'^T_H$} & \colhead{$F'^T_K$} & 
	\colhead{$F'^E_H$} & \colhead{$F'^E_K$} &
	\colhead{$( F'^E/F'^T )_H$} & \colhead{$( F'^E/F'^T )_K$} & 
	\colhead{$F^{E}_H$} & \colhead{$F^{E}_K$} & \colhead{$T_c$} \\
 & \colhead{(Jy)} & \colhead{(Jy)} & \colhead{(Jy)} & \colhead{(Jy)} & 
	\colhead{(\%)} & \colhead{(\%)} & \colhead{(Jy)} & \colhead{(Jy)} & \colhead{(K)}
}
\startdata
V594 Cas & \phn 1.7 $\pm$ 0.1 & \phn 3.3 $\pm$ 0.2 & \phn 1.3 $\pm$ 0.1 & \phn 3.0 $\pm$ 0.2 &
		76.2 $\pm$ 7.0 & 91.5 $\pm$ 2.7 & \phn 1.7 $\pm$ 0.2 & \phn 3.5 $\pm$ 0.2 & 1382 $\pm$ 117 \\
AB Aur	  & \phn 8.9 $\pm$ 0.4 & 10.8 $\pm$ 0.5 & \phn 6.1 $\pm$ 1.1 & \phn 8.9 $\pm$ 0.9 &
		68.7 $\pm$ 11.5 & 82.9 $\pm$ 6.5 & \phn 6.5 $\pm$ 1.2 & \phn 9.3 $\pm$ 0.9 & 1814 $\pm$ 331 \\
T Ori     & \phn 1.2 $\pm$ 0.01 & \phn 2.0 $\pm$ 0.1 & \phn 0.9 $\pm$ 0.1 & \phn 1.8 $\pm$ 0.1 &
		74.8 $\pm$ 8.6 & 89.5 $\pm$ 3.7 & \phn 1.1 $\pm$ 0.1 & \phn 2.0 $\pm$ 0.1 & 1503 $\pm$ 110 \\
V380 Ori-A & \phn 0.9 $\pm$ 0.05 & \phn 1.8 $\pm$ 0.1 & \phn 0.8 $\pm$ 0.1 & \phn 1.7 $\pm$ 0.1 &
		84.3 $\pm$ 4.3 & 94.2 $\pm$ 1.6 & \phn 1.0 $\pm$ 0.1 & \phn 1.9 $\pm$ 0.1 & 1457 $\pm$ 113 \\ 
$\omega$ Ori     & 12.6 $\pm$ 0.6 & \phn 8.1 $\pm$ 0.4 & \phn 0.1 $\pm$ 3.7 & \phn 0.5 $\pm$ 2.4 &
	     \phn 0.9 $\pm$ 29.8 & \phn 5.9 $\pm$ 29.1 & \phn 0.1 $\pm$ 3.8 & \phn 0.5 $\pm$ 2.4 & \nodata \\
MWC 147-A   & \phn 2.1 $\pm$ 0.1 & \phn 3.3 $\pm$ 0.2 & \phn 1.3 $\pm$ 0.2 & \phn 2.8 $\pm$ 0.2 &
		62.8 $\pm$ 9.2 & 84.3 $\pm$ 3.9 & \phn 1.7 $\pm$ 0.3 &  \phn 3.2 $\pm$ 0.2 & 1467 $\pm$ 189 \\
MWC 166-A   & \phn 4.7 $\pm$ 0.2 & \phn 4.2 $\pm$ 0.2 & \phn 0.5 $\pm$ 0.9 & \phn 1.3 $\pm$ 0.7 &
		11.7 $\pm$ 20.1 & 31.2 $\pm$ 15.8 & \phn 0.8 $\pm$ 1.4 & \phn 1.6 $\pm$ 0.8 & 1408 $\pm$ 1690 \\
MWC 863-A   & \phn 2.6 $\pm$ 0.1 & \phn 3.6 $\pm$ 0.2 & \phn 1.2 $\pm$ 0.5 & \phn 2.6 $\pm$ 0.4 &
		47.8 $\pm$ 18.2 & 73.1 $\pm$ 9.8 & \phn 1.6 $\pm$ 0.6 & \phn 3.0 $\pm$ 0.4 & 1471 $\pm$ 405 \\
MWC 275   & \phn 5.8 $\pm$ 0.3 & \phn 8.0 $\pm$ 0.4 & \phn 3.0 $\pm$ 1.1 & \phn 6.2 $\pm$ 0.8 &
		51.1 $\pm$ 19.0 & 77.0 $\pm$ 9.3 & \phn 3.1 $\pm$ 1.2 & \phn 6.3 $\pm$ 0.9 & 1393 $\pm$ 375 \\
MWC 297   & 14.4 $\pm$ 0.7 & 36.3 $\pm$ 1.8 & \phn 8.2 $\pm$ 1.4 & 30.6 $\pm$ 2.1 &
		56.6 $\pm$ 8.5 & 84.1 $\pm$ 3.1 & 23.8 $\pm$ 4.2 & 56.3 $\pm$ 4.1 & 1268 $\pm$ 143 \\
MWC 614   & \phn 2.0 $\pm$ 0.1 & \phn 2.7 $\pm$ 0.1 & \phn 0.9 $\pm$ 0.4 & \phn 1.9 $\pm$ 0.3 &
		44.7 $\pm$ 17.8 & 71.4 $\pm$ 9.6 & \phn 1.1 $\pm$ 0.4 & \phn 2.1 $\pm$ 0.3 & 1453 $\pm$ 387 \\
V1295 Aql & \phn 2.3 $\pm$ 0.1 & \phn 3.0 $\pm$ 0.1 & \phn 1.0 $\pm$ 0.5 & \phn 2.2 $\pm$ 0.3 &
		44.0 $\pm$ 19.8 & 72.0 $\pm$ 10.3 & \phn 1.0 $\pm$ 0.5 & \phn 2.2 $\pm$ 0.3 & 1325 $\pm$ 424 \\
V1685 Cyg & \phn 1.7 $\pm$ 0.1 & \phn 3.0 $\pm$ 0.1 & \phn 1.2 $\pm$ 0.1 & \phn 2.7 $\pm$ 0.2 &
		70.6 $\pm$ 7.2 & 88.4 $\pm$ 2.9 & \phn 1.8 $\pm$ 0.2 & \phn 3.4 $\pm$ 0.2 & 1463 $\pm$ 126 \\
MWC 361-A   & \phn 6.0 $\pm$ 0.3 & \phn 8.1 $\pm$ 0.4 & \phn 2.4 $\pm$ 0.9 & \phn 5.6 $\pm$ 0.7 & 
		40.5 $\pm$ 14.8 & 69.4 $\pm$ 7.7 & \phn 3.3 $\pm$ 1.3 & \phn 6.6 $\pm$ 0.9 & 1408 $\pm$ 387 \\
MWC 1080-A  & \phn 3.6 $\pm$ 0.2 & \phn 7.3 $\pm$ 0.4 & \phn 2.9 $\pm$ 0.3 & \phn 6.6 $\pm$ 0.4 &
		79.2 $\pm$ 5.2 & 91.0 $\pm$ 2.3 & \phn 6.6 $\pm$ 0.6 & 10.4 $\pm$ 0.7 & 1673 $\pm$ 151     
\enddata
\end{deluxetable}

\clearpage

\section{MODELS OF THE BRIGHTNESS DISTRIBUTION} \label{models}

The small sizes of the central stars derived in Section~\ref{fluxes}
imply that the interferometer resolves the CS material source of the
excess emission.  In this section we make use of the interferometer
data to characterize the geometrical distribution of this CS
material. This information, together with the intrinsic excess fluxes
derived in Section~\ref{fluxes}, permits the brightness distribution
to be estimated.

	\subsection{A Source with no IR Excess: $\omega$ Ori}

\begin{figure}[htbp]
\begin{center}
\includegraphics[height=4in]{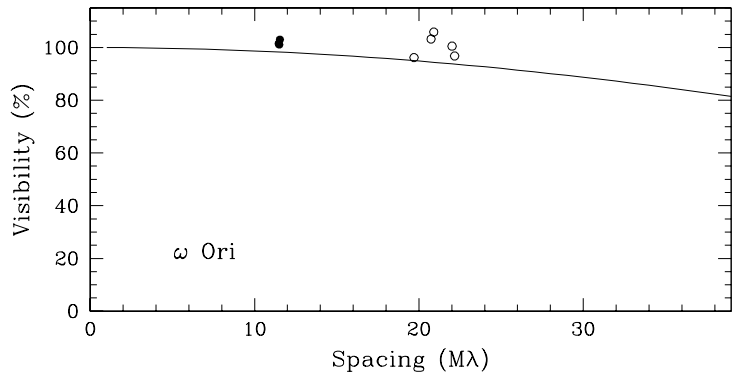}
\vspace{-1.8in}
\caption{$\omega$ Ori visibility data (H-band) and model curve
corresponding to the upper limit UD diameter ($\theta \leq 1.3$ mas).
\label{unres2}}
\end{center}
\end{figure}

The visibility data of Figure~\ref{unres2} shows that $\omega$ Ori
is unresolved by the interferometer.  Moreover, it was shown in
Section~\ref{fluxes} that there is no significant IR excess associated
with this star. Consequently, we will place a size limit to the star
itself by considering a single uniform disk (UD) component.

Using the visibility data obtained at the longest 38 m baseline, which
provides the highest angular resolution, the mean fringe spacing
projected on the sky is $s = 21.1 M\lambda$ and the mean visibility is
$V = 100.5 \pm 2.0 \%$, where the error is the error in the mean of
the 5 long baseline measurements.  Therefore the minimum visibility
allowed by the data is $V_{min} = 96.5 \%$ and by solving in the
equation for the UD visibility (recall Section~\ref{mwc1080}) it is
found that the upper limit to the UD angular diameter is $\theta \leq
1.3$ mas, at the $2\sigma$ level. The data and visibility function
corresponding to this upper limit are shown in Figure~\ref{unres2}.

We note that among our sample, this is the source which most clearly
shows the absence of IR excess above photospheric levels (see
Figure~\ref{seds}), and as such was classified as Group~III by HSVK.
Therefore, observations of this source are valuable as a control
experiment, and our result is indeed consistent with it being a naked
photosphere.

	\subsection{Models of the Source of IR Excess}

For the remainder of our sample, we consider four plausible models of
the source brightness. Each model has two components, with the stellar
emission arising in a central point source and the excess near-IR
emission contributed by either: (1) a Gaussian brightness
distribution, (2) a uniformly bright ring, (3) an infrared companion
or (4) a ``classical'' accretion disk with a temperature law $T(r)
\propto r^{-3/4}$. The Gaussian and ring models are intended to
provide a characteristic size for the source of near-IR excess, and
consistent with the observation that the visibilities remain constant
with projected baseline orientation, we consider the simplest case
that they appear circularly symmetric on the sky.

In general, the source specific intensity (i.e. brightness) is given
by

\be
I_{\nu}(\xi,\eta)  = 
	F^{\star}_{\nu} \cdot \delta(0,0) + I^E_{\nu}(\xi,\eta) \\
\ee

\noindent and the corresponding normalized visibility amplitude 
is calculated by application of the Van Cittert-Zernike
theorem (see e.g. \citealt{Goodman:1985})

\be
V_{\nu}(u,v)  =  \left| \frac{\mbox{FT}\{I_{\nu}(\xi,\eta)\}}%
	{F^{\star}_{\nu}+F^{E}_{\nu}} \right| 
	 =  \left| \frac{ F^{\star}_{\nu} +%
	 \mbox{FT}\{I^E_{\nu}(\xi,\eta)\}}%
	{F^{\star}_{\nu}+F^{E}_{\nu}} \right|
\ee

\noindent where in the above equations, $(\xi,\eta)$ are angular
offsets with respect to the source center along the projected baseline
coordinates $(u,v)$ measured in wavelength units (see e.g.
\citealt*{Thompson:1986}), and $\mbox{FT}$ denotes the Fourier
transform\footnote{Note that in the above and following visibility
equations, and in order to simplify the notation, we use intrinsic
rather than apparent fluxes, due to the fact that under our assumption
that the central star and excess source suffer the same extinction the
apparent and intrinsic fluxes relative to the total are the same. The
sizes derived, however, do not depend on this assumption.}.

In addition to $\omega$~Ori, we find 3 more sources which appear
unresolved, namely V594~Cas, T~Ori and MWC~147. However, since in
those cases there is significant near-IR excess (see
Figure~\ref{seds}), we will place appropriate upper limits to the
extended component within the context of the models
considered.  The source MWC~166 appears to have low visibility
($V=88.2 \pm 0.9$ \%).  However, it was found in the previous chapter
to have a marginally significant amount of near-IR excess flux, with
$88 \pm 20$ \% of the total H flux arising in the central star. The
data is therefore consistent with this source having an extended
component which is completely resolved by the interferometer. Given
the large error in the derived excess flux however, we cannot rule out
the possibility that there is zero excess, so that all the emission
comes from the star alone, or that the interferometer only partially
resolves an extended component which contributes a larger fraction (up
to 32\%) of the total H flux. In this case then, lower limits will be
given for the size of the excess region under the models mentioned
above.

	\subsubsection{Gaussian Intensity}
	\label{gausssection}

We consider a circularly symmetric brightness distribution given by

\begin{eqnarray}
I^E_{\nu}(\xi,\eta) & = & 
	I^0_{\nu} \cdot e^{\frac{-4 \ln 2(\xi^2+\eta^2)}{\theta_G^2}} \\
\noalign{\hbox{so that}}
V_{\nu}(s) & = & \left| \frac{ F^{\star}_{\nu} + F^E_{\nu} \cdot 
e^{\frac{-\pi^2 s^2 \theta_G^2}{4 \ln 2}}}{F^{\star}_{\nu} + F^E_{\nu}} \right|
\label{vg}
\end{eqnarray}

\noindent where $I^0_{\nu}$ is the peak intensity and $\theta_G$ is the
angular FWHM in both sky coordinates ($\xi$ and $\eta$).

Once the Gaussian FWHM is found by fitting Equation~\ref{vg} to 
the visibility data, the peak intensity is obtained from the constraint
that the flux in this component matches the measured excess flux

\be
F^E_{\nu} = \int \int I^0_{\nu} \cdot 
	e^{\frac{-4 \ln 2(\xi^2+\eta^2)}{\theta_G^2}} d\xi d\eta
        = I^0_{\nu} \cdot \Omega_G \nonumber
\ee

\noindent where $\Omega_G = \pi^2 \theta_G^2/4 \ln 2$, is the
effective solid angle. In practice, we find it most informative to
express the peak intensity in terms of the equivalent peak brightness
temperature, defined by $I^0_{\nu} = B_{\nu}(T^0_B)$. The brightness
temperature can then be readily compared with the color temperature
derived from the H and K excess fluxes in order to determine whether
this model is consistent with blackbody emission at the color 
temperature of the H and K excess.

Table~\ref{gausstable} summarizes the results obtained. In addition to
the best fitting Gaussian FWHM, the table includes the RMS of the fit
evaluated as $\sqrt{\chi^2_{min}/(N-1)}$ with $\chi^2 = \sum_{i=1}^N (
V^{data}_i - V^{model}_i )^2$, the number of individual observations
for each star (N), the linear diameter (D) corresponding to the
measured angular diameter calculated using the distances of
Table~\ref{targetstable}, and the peak brightness temperature
($T^0_B$) calculated as described above. For the two stars which were
observed in both the H and K$'$ bands and are resolved (AB~Aur and
MWC~863-A), it is found that the sizes in this model are, within the
errors, the same, and the table shows in those cases the size that is
derived from a simultaneous fit to both data sets. For the other
stars, for which an independent measurement of the size in the
K$'$-band is not available, the peak brightness temperature at K$'$ is
calculated using the size derived from fitting the H-band data.  In
all cases (as well as in the following tables corresponding the other
models under consideration), the errors in the angular sizes reflect
the uncertainty in the relative flux in each component, given in
Table~\ref{excesstable}, rather than the much smaller formal error (of
order 0.1 mas) associated with fitting a given set of visibility data.
Figure~\ref{gaussringfig} shows the data and models (solid lines)
computed with the parameters of Table~\ref{gausstable}.

We note that for AB~Aur, a slightly better agreement is obtained
by fitting the H and K$'$-band data sets separately. In that case we
derive: $\theta_G^H\,=\,4.1 \pm 0.5$ mas and $\theta_G^{K'}\,=\,4.6
\pm 0.2$ mas, with fit RMS of 3.4\% and 5.9\% respectively, to be
compared to 3.5\% and 9.6\% obtained by fitting the model of
Table~\ref{gausstable} to the H and K$'$ data separately. However,
although formally a better fit, the different H and K$'$-band sizes
are within their errors. In \citet{Millan:1999b}, based on a smaller
data set and using a different data reduction method, the opposite
result was obtained, that the characteristic size for AB Aur appeared
somewhat smaller in the K$'$ band than in the H band; although the
difference was again judged to be most likely insignificant. This
important question will need to be settled through improved accuracy
of our visibility calibration.


\begin{deluxetable}{lcccccc}
\tablecaption{Gaussian model parameters. \label{gausstable}}
\tablecolumns{7}
\tablewidth{0pt}
\tablehead{ 
\colhead{Name} & \colhead{$\theta_G$} & \colhead{Fit RMS} &
\colhead{N} & \colhead{D} & \colhead{H-band $T^0_B$} & 
\colhead{K$'$-band $T^0_B$} \\
& \colhead{(mas)} & \colhead{(\%)} & & \colhead{(AU)} &
\colhead{(K)} & \colhead{(K)}
}
\startdata
V594 Cas	& $\leq$ 0.8		& \nodata & 6  & $\leq$ 0.5		& 
	$\geq$ 1938 & $\geq$ $2212$\tablenotemark{b}  \\
AB Aur		& 4.1 $\pm$ 0.5	        & 4.6	  & 47 & 0.6 $\pm$ 0.1		& 
	1359 $\pm$ 65 \phn  & 1258 $\pm$ 63 \phn \\
T Ori		& $\leq$ 1.4		& \nodata & 2  & $\leq$ 0.6	 	& 
	$\geq$ 1443 & $\geq$ $1424$\tablenotemark{b} \\
V380 Ori-A	& 2.4 $\pm$ 0.1         & 2.0     & 8  & 1.1 $\pm$ 0.05	 	& 
	1208 $\pm$ 22 \phn  & 1144 $\pm$ $20$\tablenotemark{b} \phn \\	
MWC 147-A	& $\leq$ 1.0		& \nodata & 9  & $\leq$ 0.8	 	& 
	$\geq$ 1765 & $\geq$ 1884 \\
MWC 166-A	& $\geq$ 4.5 & \nodata & 8 & $\geq$ 5.2 & $\leq$ 1003 & $\leq$ $916$\tablenotemark{b} \\
MWC 863-A	& 7.6 $\pm$ 1.7	        & 6.7	  & 18 & 1.1 $\pm$ 0.3	 	& 
	\phn 963 $\pm$ 44 \phn & \phn 865 $\pm$ 21 \phn \\
MWC 275		& $4.6^{+1.7}_{-0.8}$	& 2.8	  & 9  & $0.6^{+0.2}_{-0.1}$	& 
	1175 $\pm$ 104 & 1124 $\pm$ 104\tablenotemark{b} \\
MWC 297		& $5.8^{+0.9}_{-0.5}$	& 2.0	  & 10 & $2.6^{+0.4}_{-0.2}$	& 
	1500 $\pm$ 77 \phn & 1594 $\pm$ $92$\tablenotemark{b} \phn	\\
MWC 614		& $7.6_{-2.5}$	        & 2.0	  & 13 & $1.8_{-0.6}$	 	& 
	\phn 923 $\pm$ 74 \phn	 & \phn 826 $\pm$ $70$\tablenotemark{b} \phn \\
V1295 Aql	& $5.4_{-1.4}$	        & 4.3	  & 4  & \nodata\tablenotemark{a}	 	& 
	\phn 991 $\pm$ 78 \phn	 & \phn 910 $\pm$ $68$\tablenotemark{b} \phn \\
V1685 Cyg	& 2.6 $\pm$ 0.1         & 2.8	  & 3  & 2.5 $\pm$ 0.1	 	& 
	1284 $\pm$ 30 \phn & 1235 $\pm$ $30$\tablenotemark{b} \phn \\
MWC 1080-A	& 2.7 $\pm$ 0.2         & 6.4	  & 20 & 5.9 $\pm$ 0.2	 	& 
	1566 $\pm$ 33 \phn & 1535 $\pm$ 35 \phn \\
\enddata
\tablenotetext{a}{No distance information available} 
\tablenotetext{b}{Calculated using the size derived from H-band visibility data}
\end{deluxetable}

By comparing the peak brightness temperatures of
Table~\ref{gausstable} with the color temperatures of
Table~\ref{excesstable}, it is apparent that for most objects, the
brightness temperature is lower than the color temperature.  One
possible physical scenario for the Gaussian model is as an
approximation to the brightness resulting from a circumstellar
environment in which the dust particles are distributed in an envelope
around the central star.  A possible interpretation then is that the
near-IR emission is optically thin, in which case the observed color
temperature depends on the physical temperature of the dust grains,
their emission properties and the optical depth of the CS cloud.  We
now further explore the consequences of this observation under the
simplifiying assumption that a single temperature characterizes the
dust envelope.  If the near-IR frequency dependence of the optical
depth in the circumstellar environment of HAEBE stars follows the
properties characteristic of interstellar dust ($A_V \sim
\lambda^{-1.9}$, due to dust particles smaller than the wavelength of
light), we may derive the physical temperature (T) of the grains and
the peak optical depth by solving in

\be
\mbox{\bf Case A:\hspace{0.1in}} \left\{ \begin{array}{lll}
B_H(T^0_B) & = & B_H(T) \cdot (1-e^{-\tau^0_H}) \\
B_K(T^0_B) & = & B_K(T) \cdot (1-e^{-\tau^0_K}) \\
\tau^0_H/\tau^0_K & = & A_H/A_K = 1.75
\end{array}
\right.
\ee

\noindent where the ratio of H to K-band extinction has again been
taken from \citet*{Steenman:1991} with $R_V=3.1$.

On the other hand, from studies of the stellar extinction toward HAEBE
stars, some authors find evidence for anomalous extinction laws ($R_V
\simeq 5$) indicative of the presence of a significant population of
grains with sizes greater (radius $a \sim 0.2 \, \mu$m) than the
average in the interstellar medium \citep*{Tjin:1978,Pezzuto:1997}.
In the limit that the dust grains become larger than the wavelength of
light, the cross-section for absorption and scattering becomes
independent of wavelength and results in neutral extinction (as in
interplanetary dust grains, of $a \sim 1-100\,\mu m$). Therefore, we
also compute the peak optical depth that results from making the
assumption that the opacity is wavelength independent

\be
\mbox{\bf Case B:\hspace{0.1in}}
B_{\nu}(T^0_B) = B_{\nu}(T_C) \cdot \left( 1-e^{-\tau^0} \right) 
\ee

The results obtained under these two assumptions are summarized in
Table~\ref{tautable}. The ``Case A'' assumption results in an estimate
of the physical temperature (T) of the dust grains and H-band peak
optical depth, as well as the visual extinction of the star due to the
CS cloud that would be implied, calculated according to $A_V =
A_H/0.14 = (\tau_H/2) \cdot (2.5/0.14\ln10)$.  For the ``Case B''
assumption, the physical temperature equals the color temperature of
Table~\ref{excesstable}, and we calculate the frequency independent
peak optical depth ($\tau^0$) and the corresponding visual extinction,
$A_V = (\tau/2) \cdot (2.5/\ln10)$.  Finally, for comparison, the
table reproduces in the last column the measured value of the visual
extinction toward the star obtained from the literature (see
Table~\ref{targetstable}).


\begin{deluxetable}{lcccccc}
\tablecaption{Gaussian model temperature and peak optical depth. 
\label{tautable}}
\tablecolumns{7}
\tablewidth{0pt}
\tablehead{ 
 & \multicolumn{3}{c}{\underline{Case A}} & 
\multicolumn{2}{c}{\underline{Case B}} & 
\colhead{\underline{Measured}} \\
\colhead{Name} & \colhead{T} & \colhead{$\tau_H^0$} & \colhead{$A_V$} &
\colhead{$\tau^0$} & \colhead{$A_V$} & \colhead{$A_V$} \\ 
& \colhead{(K)} & & & \colhead{($T=T_C$)} & & 
}
\startdata
AB Aur		& 1423 $\pm$ 61  & 1.4 $\pm$ 0.1 & 5.3 & \phn 0.2 $\pm$ 0.1 \phn  & 0.1	& 0.5 \\
V380 Ori-A	& 1238 $\pm$ 23  & 1.8 $\pm$ 0.1 & 5.8 & \phn 0.3 $\pm$ 0.05  & 0.2	& 1.43 \\
MWC 863-A	& 1104 $\pm$ 110 & 0.4 $\pm$ 0.2 & 1.5 & 0.04 $\pm$ 0.02 & 0.02 & 1.61 \\
MWC 275		& 1195 $\pm$ 100 & 2.1 $\pm$ 0.3 & 8.2 & \phn 0.4 $\pm$ 0.2 \phn   & 0.2  & 0.25 \\
MWC 297		& \nodata\tablenotemark{a} & \nodata & \nodata &  \nodata\tablenotemark{a} & \nodata		  & 8.3 \\
MWC 614		& 1074 $\pm$ 58	 & 0.3 $\pm$ 0.1 & 1.2 & 0.03 $\pm$ 0.01 & 0.02 & 1.27 \\
V1295 Aql	& 1072 $\pm$ 86  & 0.7 $\pm$ 0.1 & 2.7 & \phn 0.1 $\pm$ 0.05  & 0.06 & 0.19 \\
V1685 Cyg	& 1299 $\pm$ 30  & 2.5 $\pm$ 0.1 & 9.7 & \phn 0.6 $\pm$ 0.1 \phn  & 0.3  & 3.16 \\
MWC 1080-A	& 1570 $\pm$ 65 & 4.1 $\pm$ 0.1 & 15.9 & \phn 1.2 $\pm$ 0.1 \phn & 0.6 & 5.27 \\ 
\enddata
\tablenotetext{a}{No solution possible with $\tau^0_H \geq 0$}
\tablecomments{The optical depths and temperatures in this table are
rounded from the solutions to equations 10 and 11, the resulting
numerical loss of accuracy is however within the qouted error bars,
which result primarily from the uncertainties in the excess fluxes.}
 
\end{deluxetable}

From the results in the last two tables, we extract the following
conclusions. First, the observed fluxes and our measured sizes are
consistent with emission from dust envelopes having finite optical
thickness (i.e. $T^0_B < T_C$), in 10 out of 13 cases, even at the
position of the peak intensity.  We note that under the other extreme
assumption concerning the extinction of excess emission, i.e. that it
is zero and this material causes all of the stellar extinction, the
intrinsic excess fluxes would be given by columns 4 and 5 of
Table~\ref{excesstable}, which are lower and would lead to even
smaller values of the peak brightness and peak optical depths
inferred. The unphysical result $T^0_B > T_C$ is found for one of the
resolved sources (MWC~297) and for two of the three unresolved sources
(V594~Cas and MWC~147), inconsistently with the model assumptions, and
is most likely a consequence of the inadequacy of the assumption that
there is a single temperature throughout the envelope rather than a
radial temperature gradient. Second, the assumption that the
extinction properties in the envelope are interstellar (case A) leads
to relatively large values of the visual extinction (median $A_V=5.8$)
in the line of sight toward the central stars. Under this assumption
then, a special geometry that allows a clear line of sight to the star
itself would need to be invoked, as has been proposed by
\citet{Berrilli:1992} and \citet*{Hartmann:1993} in the context of SED
modelling. Third, under the assumption that the extinction in the
envelope is neutral (case B), the visual extinctions to the stars
(median $A_V=0.2$) are somewhat lower than the measured
values. However, the values derived are lower limits, and therefore
consistent with the observed ones, since they do not include an
interstellar contribution, or a contribution from the larger particles
that are likely to exist in a realistic distribution of sizes.

\clearpage

	\subsubsection{Uniform Ring}

In the previous section it was shown that under the Gaussian model,
the effective sizes measured by the interferometer require that the
emission be optically thin over most of the area of the source, in
order to match the observed fluxes. An alternative solution is to
consider optically thick emission at the observed color temperature,
but reduce the emitting surface area by concentrating it in a narrow
ring around the star.

For a uniform ring of intensity $I_{\nu}=B_{\nu}(T_C)$, and inner and
outer angular diameters $\theta_1$ and $\theta_2$ respectively, we
have

\begin{eqnarray}
I^E_{\nu}(\xi,\eta) & = & \left\{ \begin{array}{ll}
	B_{\nu}(T_C) & \mbox{if 
$\quad \frac{1}{2} \theta_1 \leq \sqrt{\xi^2 + \eta^2} \leq 
\frac{1}{2} \theta_2$} \\
	0  & \mbox{otherwise} \end{array} \right. \\[0.1in]
V_{\nu}(s) & = & \left| \frac{ F^{\star}_{\nu}}{F^T_{\nu}} + 
	\frac{F^E_{\nu}}{F^T_{\nu}} \cdot 
	\frac{4}{\pi \left( \theta_2^2 - \theta_1^2 \right)}
	\cdot \left[ \frac{\theta_2 \cdot J_1(\pi \theta_2 s)}{2 s} -
	\frac{\theta_1 \cdot J_1(\pi \theta_1 s)}{2 s} \right] \right| 
\end{eqnarray}

\noindent where for each value of the inner diameter, the outer
diameter is constrained by the requirement that the flux from the ring
matches the excess flux: $\theta_2 = \sqrt{4 \Omega/\pi + \theta_1^2}$
with $\Omega = F^E_{\nu}/B_{\nu}(T_C)$. The only parameter to be fit
to the visibility data is then the ring inner diameter.

\begin{figure}[htbp]
\begin{center}
\includegraphics[height=8in]{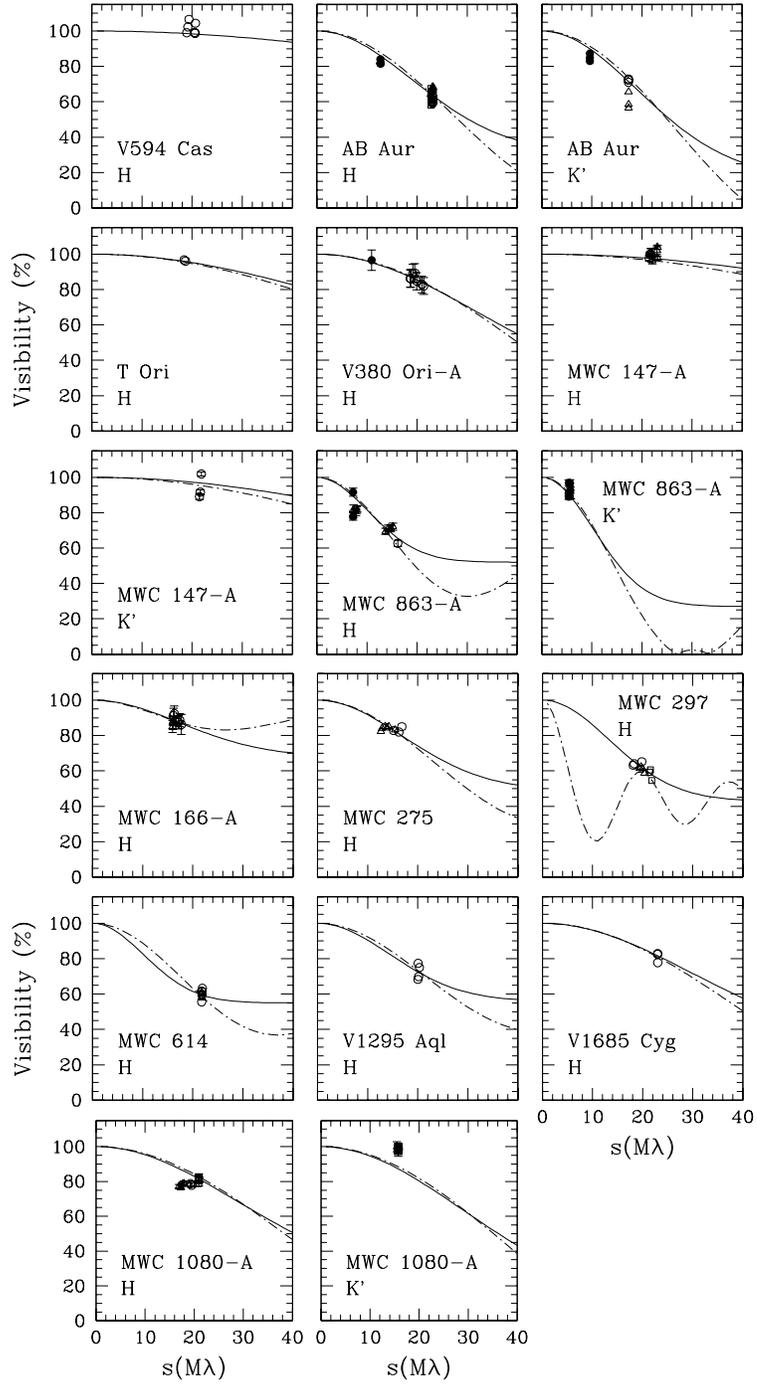}
\vspace{-0.5in}
\caption{Visibility data and Gaussian and ring models.  Solid
lines: Point source plus Gaussian intensity model. Dot-short dashed
lines: Point source plus uniform ring model. The models are calculated
for the best fit parameters of Tables~\ref{gausstable}~and~\ref{ringtable}.
\label{gaussringfig}}
\end{center}
\end{figure}

The results are shown in Table~\ref{ringtable}, where in addition to
the quantities already defined, the last two columns contain the
linear ring inner and outer diameters.  Figure~\ref{gaussringfig}
shows the data and models (dot-short dash lines) computed with the
parameters of Table~\ref{ringtable}.  We note that the source V594~Cas
(which appears unresolved) can not be represented by this model, due
to the fact that for any inner ring diameter, the outer diameter
required to match the excess flux results in a source that would
appear resolved.  Similarly, the large excess flux for MWC 297 implies
that any solution for the ring model has a large size, much larger
than the characteristic size given by the Gaussian model. The data can
formally be well fit only at the peak of the second lobe of the ring
visibility function, and we do not consider this a likely
interpretation of the data for this source.


\begin{deluxetable}{lcccccc}
\tablecaption{Ring model parameters. \label{ringtable}}
\tablecolumns{7}
\tablewidth{0pt}
\tablehead{ 
\colhead{Name} & \colhead{$\theta_1$} & \colhead{FitRMS} & \colhead{N} & 
\colhead{$\theta_2$\tablenotemark{a}} &  \colhead{$D_1$} & \colhead{$D_2$} \\
 & \colhead{(mas)} & \colhead{(\%)} & & \colhead{(mas)} & \colhead{(AU)} & 
\colhead{(AU)}  
}
\startdata
V594 Cas	& \nodata\tablenotemark{b} & \nodata & \nodata & \nodata & \nodata & \nodata  \\
AB Aur 		& 4.2 $\pm$ 0.4\phn  & 5.2 & 47 & 4.7 $\pm$ 0.4\phn & 0.6\phn $\pm$ 0.05 & 0.7 $\pm$ 0.05 \\ 
T Ori		& $\leq$ 1.4\phn & \nodata & 2 & $\leq$ 2.05 & $\leq$ 0.6 & $\leq$ 0.9 \\
V380 Ori-A	& 2.5 $\pm$ 0.1\phn  & 2.0 & 8  & 3.0 $\pm$ 0.1\phn & 1.15 $\pm$ 0.50 & 1.4 $\pm$ 0.05 \\
MWC 147-A	& $\leq$ 0.4     & \nodata & 12 & $\leq$ 2.0 & $\leq$ 0.3 & $\leq$ 1.6 \\
MWC 166-A	& $\geq 9.5$	& 2.5 & 8 & $\geq 9.6$ & $\geq 10.9$ & $\geq 11.0$ \\
MWC 863-A	& $8.2^{+2.7}_{-1.5}$ & 6.2 & 18 & $8.5^{+2.7}_{-1.5}$ & $1.2^{+0.4}_{-0.2}$ & $1.3^{+0.4}_{-0.2}$ \\
MWC 275		& $4.7^{+1.7}_{-0.9}$ & 3.1 & 9 & $5.7^{+1.4}_{-0.8}$ & $0.6^{+0.2}_{-0.1}$ & $0.7^{+0.2}_{-0.1}$ \\
MWC 297\tablenotemark{c} 	& $21.9^{+1.2}$   & 4.1 & 10 & $25.0^{+1.0}$ & $9.9^{+0.5}$ & $11.25^{+0.45}$ \\
MWC 614		& 6.7 $\pm$ 1.5\phn  & 2.3 & 13 & 6.9 $\pm$ 1.4\phn & 1.6\phn $\pm$ 0.4\phn & 1.7 $\pm$ 0.3\phn \\
V1295 Aql 	& $5.4_{-1.2}$   & 4.4 & 4 & $5.8_{-1.1}$ & \nodata\tablenotemark{c} & \nodata\tablenotemark{c}  \\
V1685 Cyg	& 2.7 $\pm$ 0.15 & 2.8 & 3 & 3.4 $\pm$ 0.15 & 2.55 $\pm$ 0.15 & 3.3 $\pm$ 0.15 \\
MWC 1080-A	& 2.3 $\pm$ 0.3\phn  & 9.0 & 20 & 3.6 $\pm$ 0.3\phn & 5.1\phn $\pm$ 0.7\phn & 7.9 $\pm$ 0.7\phn 
\enddata
\tablenotetext{a}{Derived from $\theta_1$ and flux}
\tablenotetext{b}{No solution can match flux and visibility data simultaneously, see text}
\tablenotetext{c}{The only formal solution requires $\theta_1 \gg \theta_G$, see text} 
\tablenotetext{d}{No distance information is available}
\end{deluxetable}

\clearpage

	\subsubsection{Accretion Disk}

Given that the stars in our sample which the interferometer resolves
are all Group~I sources in the classification of HSVK, we consider in
this section whether such an accretion disk model is consistent with
the interferometer data.

We consider a flat, blackbody disk heated by a central star of
effective temperature $T_{\mbox{eff}}$ and radius $R_{\star}$.  Energy
balance between stellar light absorbed at distance $r$ by one side of
the disk and subsequent re-radiation at the equilibrium temperature,
results in a temperature law in the disk, for $R_{\star}/r \ll 1$
given by \citep{Friedjung:1985}

\be
T_{rep}(r) = T_{\star} \cdot \left( \frac{1}{4} \right)^{1/4} \cdot 
	\left( \frac{R_{\star}}{r} \right)^{3/4}
\ee 

The contribution to the disk temperature resulting from viscous
heating in a steady state, optically thick, self-gravitating disk,
also for $R_{\star}/r \ll 1$, may be shown to be \citep*{Lynden:1974}

\be
T_{acc}(r) = \left( \frac{3 G M_{\star} \dot{M}_{acc}}{\sigma 8 \pi} \right)^{1/4} \cdot r^{-3/4}  
\ee

\noindent where $\dot{M}_{acc}$ is the accretion rate, $\sigma$ is the
Stefan-Boltzman constant, and the other symbols have their usual
meanings. 
 
Considering the reprocessing and accretion processes simultaneously results 
in a radial temperature law given by 
$T(r) = \left[ T_{rep}^4(r) + T_{acc}^4(r) \right]^{1/4}$.
In convenient units, we have

\begin{eqnarray}
T(r) & = & \left[ 2.55\times 10^{-8} \left(\frac{R_{\star}}{R_{\sun}} \right)^3
	T_{\star}^4 + 5.26\times 10^{10} \left( \frac{M_{\star}}{M_{\sun}} \right)
	\left( \frac{\dot{M}_{acc}}{10^{-5}} \right) \right]^{1/4} \cdot 
	r_{AU}^{-3/4} \label{teq1} \\
     & = & T_{1AU} \cdot r_{AU}^{-3/4} \label{teq2}
\end{eqnarray}

Therefore, in this model, if the properties of the central
star are known, the disk temperature by stellar heating is completely
specified; and the heating due to viscous accretion depends only on
the mass accretion rate. In this paper, when fitting the SED and
visibility data, we treat $T_{1AU}$ as the parameter which sets the
disk temperature scale, independent of the details about the
heating mechanism.

Having specified a radial temperature law, the spectral energy
distribution of the star-disk system may then be calculated by adding
the disk spectrum to the stellar spectrum, where the disk spectrum
results from a sum, from a minimum radius $R_{min}=R_{hole}$ to a
maximum radius $R_{max}$, of blackbody annuli at temperatures $T(r)$.

Since for a given disk inclination the SED completely specifies the
model ($T_{1AU}$ and $R_{hole}$), the corresponding visibility curves
may be computed and compared to the data. The result is that the
models of HSVK make the prediction that these sources should appear
unresolved or very nearly unresolved to the IOTA. This conclusion may
be qualitatively understood as follows: the temperatures at the inner
edge of the disk that result from the parameters derived by HSVK are
in the range $T(R_{hole}) = 1500 - 3100 K$ (2000 K average); and
recalling that the peak of emission for a blackbody at 2000 K is
$\lambda_{max} = 0.290 \times 10^4/2000K = 1.45 \, \mu m$, it follows
that the region from which the near-IR emission arises in those models
approximately coincides with the inner edge of the disk. Using the
average inner hole size (0.3 AU) and distance (617 pc) in their sample
implies an angular radius of only 0.5 mas for the near-IR emitting
region.

\begin{figure}[htbp]
\begin{center}
\includegraphics[height=4in]{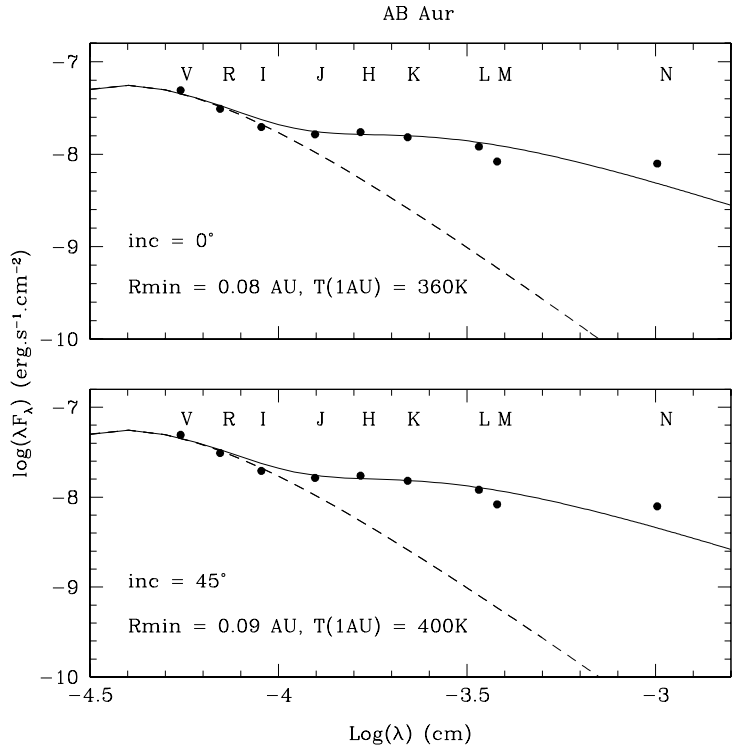}
\caption{Example of fitting the SED data to the point source plus
accretion disk model for two different values of the disk inclination,
$i=0 \arcdeg$ and $i=45 \arcdeg$.\label{hsvksedfig}}
\end{center}
\end{figure}

However, the inner hole sizes derived by HSVK are under-estimates, due
to their assumption that the disks are seen face-on. Thus, we must
consider whether a solution that matches the visibility data exists
for accretion disk models with non-zero inclination.  For every value
of the disk inclination, there will in general exist values of
$R_{hole}$ and $T_{1AU}$ such that the SED is well fit, since it is
always possible to recover the flux lost to the lower apparent surface
area by increasing the disk temperature, and to recover the
characteristic shape of the SED in the near-IR by increasing the size
of the inner hole. This degeneracy is illustrated for AB~Aur in
Figure~\ref{hsvksedfig}, which shows equivalent fits to the SED for
inclination values of $i = 0 \arcdeg$ and $i = 45 \arcdeg$ (angle
between the disk normal and the line of sight).  Having found
$R_{hole}$ and $T_{1AU}$ such that the SED is well fit for every disk
inclination between $0\arcdeg$ and $90\arcdeg$, our approach is to
determine the smallest inclination for which there exists a disk
position angle such that the visibility data is also well fit.  Note
that although a non-zero inclination results in larger sources of
elliptical shape which can reproduce the amount of resolution observed
by aligning the disk semi-major axis with the IOTA baseline, it also
predicts visibilities that vary with baseline position angle, a
feature not observed in our data set.

As before, we require that the flux in the disk component matches the 
observed excess flux. The normalized visibility 
is given by

\be 
V_{\nu}(u,v) = \left| \frac{ F^{\star}_{\nu} + \mbox{\Bd V}^{Disk}_{\nu}(u,v)}{
	F^{\star}_{\nu} + F^{E}_{\nu}} \right|;
\quad \mbox{with:} \quad 
\mbox{\Bd V}^{Disk}_{\nu}(u,v) = \mbox{FT}[I_{\nu}^{Disk}(\xi,\eta)]
\ee

The visibility function for the disk component is calculated
by summing the visibility functions of the successive annuli, which in
turn are given by the difference between the UD visibility functions
corresponding to the outer ($r_{out}$) and inner ($r_{in}$) annulus 
radii  

\begin{eqnarray}
\mbox{\Bd V}^{Disk}_{\nu}(u,v) & = & 
	\sum_{R_{min}}^{R_{max}} \mbox{\Bd V}^{Annulus}_{\nu}(u,v) \\
	& = & \sum_{R_{min}}^{R_{max}} 
		\left( \mbox{\Bd V}^{UD}_{\nu}(u,v;r_{out}) -  
		\mbox{\Bd V}^{UD}_{\nu}(u,v;r_{in})
		\right) 
\label{diskvis}
\end{eqnarray}

For a circularly symmetric UD intensity that is thus tilted by angle
$i$ and rotated so that its semi-major axis has a position angle
$\Psi$ (measured from North toward East), the visibility function is
obtained from the familiar face-on result by a coordinate
transformation and use of the scaling property of the Fourier
transform, so that each disk visibility in Equation~\ref{diskvis} is
given by

\begin{eqnarray}
\mbox{\Bd V}^{UD}_{\nu}(u,v;r)  & = & \frac{1}{2} 
	\left[B_{\nu}(T_{in}) + B_{\nu}(T_{out}) \right] \cdot
	\frac{\pi r^2}{d^2} \cdot \cos(i) 
	\cdot 2 \cdot \frac{J_1(\pi \beta)}{\pi \beta} \\
\noalign{\mbox{with}}
\beta & = & s \cdot \theta \cdot 
	\sqrt{\cos^2(\Psi - \Phi) + \cos^2(i) \sin^2(\Psi - \Phi)}
\end{eqnarray}

\noindent where the angular diameter of each UD is $\theta = 2r/d$.

\begin{figure}[htbp]
\begin{center}
\includegraphics[scale=0.8]{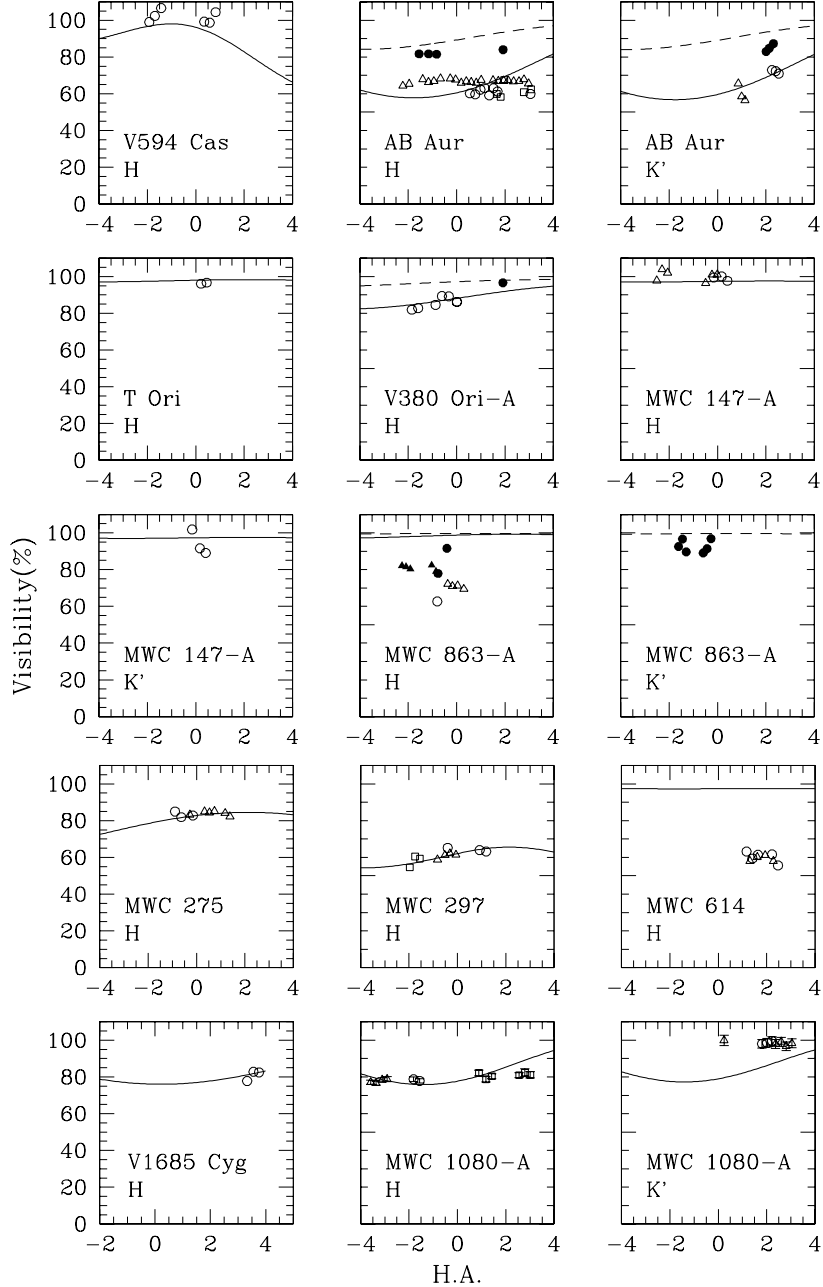}
\vspace{-1.2in}
\caption{Visibility data and point source plus accretion disk
models.  Open symbols and solid lines: long baseline data and
models. Filled symbols and dashed lines: short baseline data and
models. The models are calculated for the parameters of
Table~\ref{accretiontable}. The data and models are plotted as a
function of source hour angle (H.A.), which determines the orientation
of the projected baseline.
\label{accretionfig}}
\end{center}
\end{figure}

The results of fitting the SED and visibility data to this model are
summarized in Table~\ref{accretiontable} and
Figure~\ref{accretionfig}. In addition to the quantities already
defined, the table lists in the last column the temperatures at the
inner edge of the disk that result from Equation~\ref{teq2}.  For
those cases for which a good fit to the visibility data was found, the
best-fit parameters are listed in the table and that is the solution
plotted in the figure. Note however that for AB~Aur and MWC~1080-A,
although there is no accretion disk model which fits the visibility
data well, the data covers a relatively large range of baseline
position angles (45 and $65 \arcdeg$ respectively) and we have plotted
the closest fit that can be obtained in order to illustrate the fact
that these solutions are rejected on the basis that the data does not
show the variation with baseline position angle that is expected from
the source assymetry. For two other sources (MWC~863-A and MWC~614)
the model can not reproduce the observed visibilities for any $i <
90\arcdeg$, but as an illustration, the table parameters and
visibility curve plotted correspond to a disk inclination of $30
\arcdeg$, the expected average for a randomly selected
sample. Finally, we note that for V594~Cas, as was the case of the
ring model, due to the large near-IR excess the accretion disk model
can not reproduce the observation that the source appears
unresolved. This source is the exception, in that the closest fit
found corresponds to orienting a highly tilted disk with its {\it
small} axis along the projected baseline.

In summary, of 11 candidates, 6 can be fit with the disk model. Of
those, 2 are unresolved (T~Ori and MWC~147-A) and consistent with a
face-on disk.  Of the 4 resolved sources however, 3 require extreme
values of the disk inclination ($i \geq 80 \arcdeg$).  Based on this
statistically unlikely requirement, we consider the accretion disk a
less attractive explanation for the observed emission.

For the 4 resolved stars for which the visibility data could be fit,
we have computed in Table~\ref{accretiontab2} the temperature at 1 AU
that would result from stellar heating only ($T_{1AU}^{rep}$), by
setting $\dot{M} = 0$ in Equation~\ref{teq1}, and where we have used
our adopted stellar temperatures, the stellar radii derived from the
solid angle fits to the SEDs (Table~\ref{starfluxes}) and adopted
distances (Table~\ref{targetstable}).  Comparing with the values of
$T_{1AU}$ required to match the SED and visibility data, it can be
seen that all objects except MWC~297 require an additional source of
heating. Under the assumption that this additional heating mechanism
is viscous accretion, the table shows the value of $\dot{M}_{acc}$
that would be required, again using Equation~\ref{teq1}.  The results
reproduce those of HSVK for MWC~297, since that is the only object
which could be fit with a face-on disk (also, non-black grains are
required for this object, since we obtained $T_{1AU} <
T^{rep}_{1AU}$).  For the other sources, our implied accretion rates
are higher than predicted by HSVK, by factors of 30.1, 11.5 and 9.1
for V380~Ori-A, MWC~275 and V1685~Cyg respectively.

\vspace{0.5in}

\begin{deluxetable}{lccccccc}
\tablecaption{Accretion disk model parameters I. \label{accretiontable}}
\tablecolumns{8}
\tablewidth{0pt}
\tablehead{ 
\colhead{Name} & \colhead{$i$} & \colhead{$R_{hole}$} & \colhead{$T_{1AU}$} & 
\colhead{$\Psi$} &  \colhead{Good Fit} & \colhead{FitRMS} & 
\colhead{$T(R_{hole})$} \\
 & \colhead{($\arcdeg$)} & \colhead{(AU)} & \colhead{(K)} & 
\colhead{($\arcdeg$)} & \colhead{to Visibility Data?}  & \colhead{(\%)} & 
\colhead{(K)}  
}
\startdata
V594 Cas	& 80\tablenotemark{a} & 1.50 & 2131 & -60 & No\tablenotemark{c}  & 8.1 & 1572 \\
AB Aur		& 82\tablenotemark{a} & 0.25 & 790  & 40 & No & 10.9 & 2234  \\
T Ori		& 0  & 0.17 & 540  & any & Yes\tablenotemark{c}	 & 0.5 & 2040 \\
V380 Ori-A	& 80 & 0.41 & 1300 & 30 & Yes & 2.5 & 2537 \\
MWC 147-A	& 0  & 0.30 & 910 & any & Yes\tablenotemark{c} & 4.1 & 2245 \\ 
MWC 863-A	& 30\tablenotemark{b} & 0.07 & 280  & 30 & No & 20.1 & 2057 \\
MWC 275		& 85 & 0.24 & 720  & 0  & Yes & 2.1 & 2100 \\
MWC 297		& 0  & 0.90 & 1830 & any & Yes & 2.5 & 1980 \\
MWC 614		& 30\tablenotemark{b} & 0.12 & 380  & 0  & No & 40.2 & 1864 \\
V1685 Cyg 	& 85 & 1.05 & 2500 & 20 & Yes & 2.2 & 2410 \\
MWC 1080-A	& 80\tablenotemark{a} & 2.70 & 5350 & 50 & No &  8.5 & 2539 
\enddata
\tablenotetext{a}{No good fit exists, 
value of inclination chosen as an illustration as that which predicts 
approximately the correct size, but is ruled out due to resulting asymmetry, 
see text}
\tablenotetext{b}{No good fit exists for any value of the inclination, 
value of inclination chosen as an illustration corresponds to expected 
average for randomly selected sample, see text}
\tablenotetext{c}{Unresolved}
\end{deluxetable}

\begin{deluxetable}{lccccc}
\tablecaption{Accretion disk model parameters II. \label{accretiontab2}}
\tablecolumns{6}
\tablewidth{0pt}
\tablehead{ 
\colhead{Name} & \colhead{$T_{\star}$} & \colhead{$R_{\star}$} & 
\colhead{$T_{1AU}^{rep}$} & \colhead{$M_{\star}$} & 
\colhead{$\dot{M}_{acc}$} \\ 
 & \colhead{(K)} & \colhead{($R_{\sun}$)} & \colhead{(K)} & 
\colhead{($M_{\sun}$)} & \colhead{($10^{-5} M_{\sun} yr^{-1}$)}
} 
\startdata
V380 Ori-A	& 12190 & 1.7 & 229 & \phn 3.6 & 15.1 \\
MWC 275		& 9332  & 2.2 & 213 & \phn 3.3 & \phn 1.5 \\
MWC 297		& 33884 & 8.8 & 2188 & 26.5 & \nodata\tablenotemark{a} \\
V1685 Cyg 	& 21878 & 4.9 & 910 & 12.6 & 57.9
\enddata
\tablerefs{Stellar masses from \citet{Hillenbrand:1992}.}
\tablenotetext{a}{No accretion required} 
\end{deluxetable}

\clearpage

	\subsubsection{Infrared Companion}

The potential of an infrared companion for reproducing the observed
SEDs of HAEBE stars was recognized by \citet{Hartmann:1993}. In this
scenario, the companion is embedded in a dust envelope and appears as
an infrared source.  Although the only star for which a clear binary
detection has been made is MWC 361-A, we must consider the possibility
that the other stars appear resolved because we have sampled only a
few points of what is really the sinusoidal visibility curve of a
binary system\footnote{\citet*{Corporon:1999} also surveyed 42 HAEBE
stars spectroscopically and detected 14 companions; 7 through Li I
detection and 6 through radial velocity variations.  For those stars
which are adaptive optics binaries, the large separations measured
imply that those companions (the ``B'' components) are unlikely to be
responsible for the radial velocity variations detected, and that a
third star may be involved instead. Since we measure the fringe
corresponding to the ``A'' adaptive optics components only, it is
plausible that it is the spectroscopic companions which are
responsible for the low visibilities we observe.}.

In general, the normalized visibility of the central fringe (zero OPD)
for a binary system where the companion is at offset $(\xi_2,\eta_2)$
is given by

\be
V^2_{\nu}(u,v) =  A_{\nu}^2 + B_{\nu}^2 \cdot \mbox{F}^2(\Delta w) + 
	2 \cdot A_{\nu} \cdot B_{\nu} \cdot \mbox{F}(\Delta w) \cdot 
	\cos[2 \pi (u \xi_2 + v \eta_2)] 
\ee

\noindent where the ``fringe washing function'' $\mbox{F}(\Delta w) =
\sin(\pi \Delta \nu \Delta w)/(\pi \Delta \nu \Delta w)$ accounts for
the reduction in amplitude due to, for a given separation in delay
$\Delta w (\mu m)= \lambda_0(-u \xi_2 + v \eta_2)$, the finite
bandwidth of the spectral filter (represented here by a square
function of wavenumber width $\Delta \nu$).  The amplitude parameters
in the above equation are the product of the fraction of the total
flux in each component times their normalized visibility as individual
sources

\begin{eqnarray}
A_{\nu} & = & \frac{F_{\nu,1}}{F_{\nu,1}+F_{\nu,2}} \cdot V_{\nu,1} \label{A} \\
B_{\nu} & = & \frac{F_{\nu,2}}{F_{\nu,1}+F_{\nu,2}} \cdot V_{\nu,2} \label{B}
\end{eqnarray}

Therefore, there are four parameters in the model: $\xi_2,\eta_2, A_{\nu}$
and $B_{\nu}$, and unless the data provides good coverage in baseline
coordinates, further assumptions are needed in order to properly
constrain it.  Thus, we have divided our fits to binary models
into three categories.

	\subsubsubsection{Binary model I. Projected separations}

\begin{figure}[htbp]
\begin{center}
\includegraphics[height=4in]{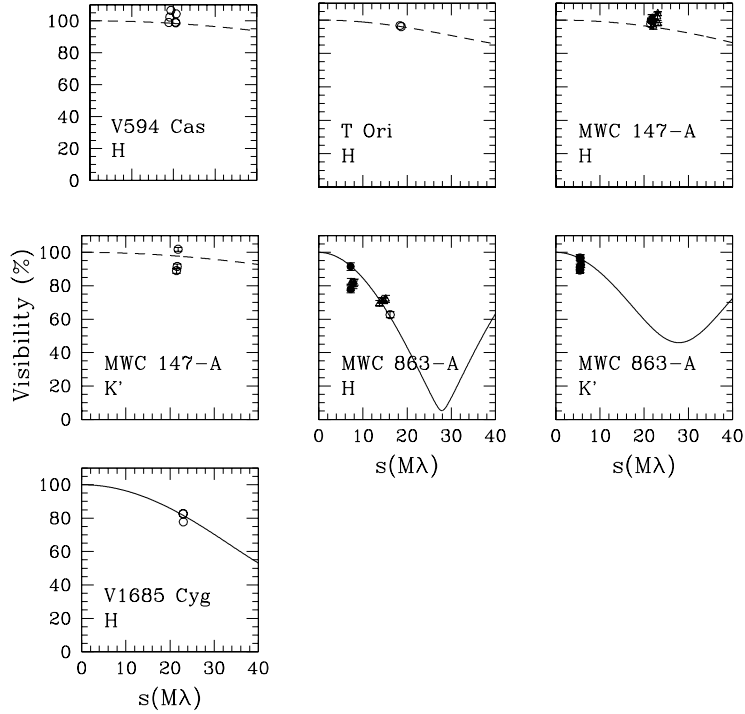}
\caption{Binary model I: Projected binary separations. The dashed
lines are upper limits for the unresolved sources.  The models are
calculated for the best fit parameters of Table~\ref{bintab1}.  
\label{bin1dfig}}
\end{center}
\end{figure}

In the case that the data are obtained under essentially a constant
baseline position angle, at each epoch, we find the angular separation
$\theta$ between the binary components ``projected'' in the direction
of the baseline vector defined by $(u \xi_2 + v \eta_2) = s \cdot
\Theta \cdot \cos(\Phi-\Upsilon) = s \cdot \theta \nonumber$; where
$\Theta$ is the true angular separation and $\Upsilon$ is the position
angle of the binary vector (measured from North, toward East).
Furthermore, we assume that each star is unresolved ($V^{\nu}_{1,2} =
1.0$), and that all of the near-IR excess is due to the companion
star, so that the amplitude of the sinusoidal visibility is given by
the flux ratio, $F_{\nu,2}/F_{\nu,1}=F^E_{\nu}/F^{\star}_{\nu}$, of
the binary components.  With these assumptions, there is one parameter
left to fit, namely the projected angular separation $\theta$.  These
assumptions are also used in the case of the unresolved sources in
order to determine a limit to this parameter.  The results are
summarized in Table~\ref{bintab1} and Figure~\ref{bin1dfig}.

\begin{deluxetable}{lcccc}
\tablecaption{Binary model parameters I. \label{bintab1}}
\tablecolumns{5}
\tablewidth{0pt}
\tablehead{
\colhead{Name} & \colhead{$(F_2/F_1)_{H;K}$} & 
\colhead{$\theta$} & \colhead{Fit RMS} & 
\colhead{N} \\
 &  & \colhead{(mas)} & \colhead{(\%)} &
}
\startdata
V594 Cas	& 3.3 $\pm$ 0.9 & $\leq$ 0.7 & \nodata & 6 \\
T Ori		& 9.0 $\pm$ 3.2 & $\leq$ 1.7 & \nodata & 2 \\
MWC 147-A	& 1.7 $\pm$ 0.5;\,5.6 $\pm$ 1.2 & $\leq$ 0.9 & \nodata & 12 \\
MWC 863-A	& 0.9 $\pm$ 0.5;\,2.7 $\pm$ 0.9 & 3.7 $\pm$ 0.3 & 7.1 & 18 \\
V1685 Cyg	& 2.4 $\pm$ 0.5 & 1.96 $\pm$ 0.1 & 2.8 & 3
\enddata
\end{deluxetable}

	\subsubsubsection{Binary model II. Limits on companion position}

For eight of the targets, the visibility data remains constant within
the errors, even as the baseline position angle changes by a
significant amount. In these cases, the lack of visibility variation
provides a useful constraint to the range of possible binary
solutions.  Without a detection of the sinusoidal visibility however,
the relative fluxes, and amount of resolution achieved in each
component are undetermined. Thus, here we also assume that the
totality of the near-IR excess is due to the companion, and that both
are unresolved.  The two parameters left to fit are then the angular
offsets $\xi_2$ and $\eta_2$ of the companion star. Because observations
made at different epochs are typically separated by several months,
during which the companion position may have changed, we establish the
possible binary solutions using the single-epoch set of data with the 
largest baseline position angle coverage.

\begin{figure}[htbp]
\begin{center}
\includegraphics[height=8in]{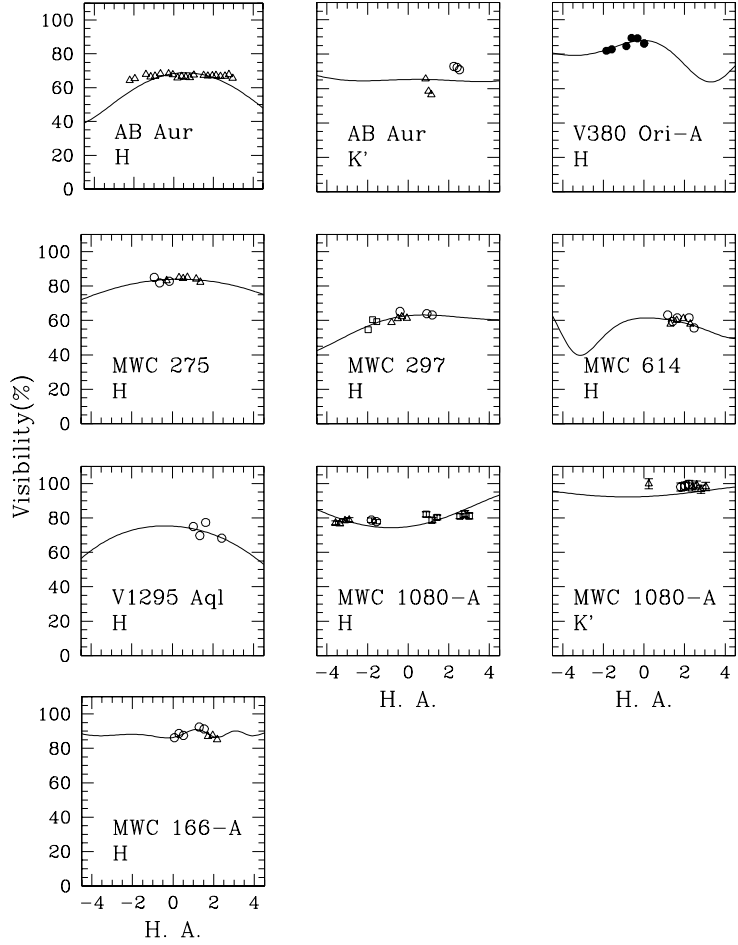}
\vspace{-1.5in}
\caption{Binary model II: Companion offsets.  The models are
calculated for the best fit parameters of Table~\ref{bintab2}.  The
data and models are plotted as a function of source hour angle (H.A.),
which determines the orientation of the projected baseline.  In these
fits only visibility data corresponding a single epoch with the
highest range of baseline rotation, which provides maximum sensitivity
to binary detection, is used.
\label{bin2dfig}}
\end{center}
\end{figure}

The results of exploring the solutions allowed by the data are
summarized in Table~\ref{bintab2}, and include, in addition to already
defined quantities, the formal best fit for the angular offsets of the
companion (columns 3 and 4) and the range in binary position angle
corresponding to other solutions consistent with the data (column 6).
The formal best fits are plotted together with the data in
Figure~\ref{bin2dfig}.

\begin{deluxetable}{lcccccc}
\tablecaption{Binary model parameters II. \label{bintab2}}
\tablecolumns{7}
\tablewidth{0pt}
\tablehead{
& & \multicolumn{4}{c}{Best Fit} & \\ \cline{3-6}
\colhead{Name} & \colhead{$(F_2/F_1)_{H;K}$} & \colhead{$\xi_2$} 
& \colhead{$\eta_2$} &
\colhead{Fit RMS} & \colhead{N} & \colhead{$\Upsilon$} \\
 &  & \colhead{(mas)} & \colhead{(mas)} & \colhead{(\%)} & & 
\colhead{($\arcdeg$)} 
}
\startdata
AB Aur	& 2.2 $\pm$ 0.8;\,4.7 $\pm$ 1.9 & 1.3 $\pm$ 0.1 & \phn 6.3 $\pm$ 0.04 & 3.4 		& 21 & 11.6 \\
V380 Ori-A	& 5.6\phn $\pm$ 1.6 & -8.4 $\pm$ 0.4 & 15.9 $\pm$ 0.4 & 1.8 & 7 & 
	$[-27.8,+12.9]$  \\
MWC 166-A	& 0.1\phn $\pm$ 0.2 & -29.9 $\pm$ 4.1 & 89.7 $\pm$ 1.1 & 1.4 & 8 &
	$[-35.0,+21.5]$ \\ 
MWC 275		& 1.04 $\pm$ 0.6 & -0.6 $\pm$ 0.2 & \phn 3.2 $\pm$ 0.1 & 1.3 & 9 & 
	$[-13.2,-11.9]$ \\
MWC 297		& 1.3\phn $\pm$ 0.3 & 1.6 $\pm$ 0.9 & 45.8 $\pm$ 0.3 & 1.9 & 10 & 
	$[-1.3\phn,+8.1\phn]$ \\
MWC 614		& 0.8\phn $\pm$ 0.4 & 4.6 $\pm$ 2.6 & 45.9 $\pm$ 0.4 & 2.2 & 9 & 
	$[-9.9\phn,+17.9]$ \\
V1295 Aql	& 0.8\phn $\pm$ 0.4 & 1.9 $\pm$ 1.9 & 28.9 $\pm$ 0.3 & 4.5 & 4 & 
	$[-7.4\phn,+26.2]$ \\
MWC 1080-A & 3.8 $\pm$ 0.1;\,10.1 $\pm$ 3.5 & 2.1 $\pm$ 0.2 & \phn 2.4 $\pm$ 0.1 & 
	3.5 & 12 & $41.2$
\enddata
\end{deluxetable}

In summary of the last two sub-sections, a binary model can be ruled
out for AB~Aur, since there is no solution that reproduces the
constant visibility observed on this source over the $45 \arcdeg$
range of baseline rotation.  For the other sources, a binary model can
potentially explain the visibility data. For the case~II sources, the
fit residuals are somewhat smaller than in the case of the Gaussian
and ring models, as one would expect since there are two free
parameters in this model. However, the difference is within the
expected calibration errors and we do not believe it to be
significant.  We note that for MWC~166-A the solution with smallest
angular offsets corresponds to a separation of $\sim 0.1 \arcsec$,
thus it is unlikely that such companion would have been un-detected in
the adaptive optics survey by \citet{Corporon:1998}.


	\subsubsubsection{Binary model III. Solution for MWC 361-A}

For this source, the visibility data shows the clear signature of a
binary system, detected interferometrically for the first
time. Observations were made at two epochs, June 1998 and September
1998; and both H and K$'$ band data were obtained in each epoch.  It
may be seen from the data that, because the maximum visibility is less
than 100\%, a solution where at least one of the components is
resolved is required. Therefore, we fit the data to the four
parameters $A_{\nu}$, $B_{\nu}$, $\xi_2$ and $\eta_2$.

\begin{figure}[htbp]
\begin{center}
\includegraphics[height=4in]{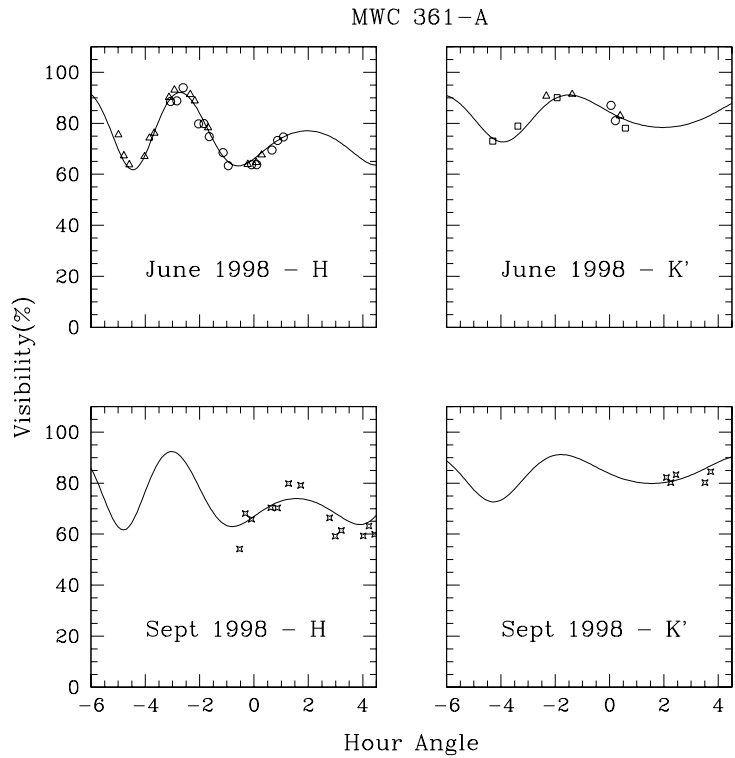}
\caption{Binary model solution for MWC 361-A.
\label{mwc361fig}}
\end{center}
\end{figure}

The four sets of data do not have the same power in constraining the
model, since they do not sample the visibility curve equally well. For
this reason, we proceed by finding the best fit solution to all four
parameters using the best set, June 1998 H-band data. The position for
the companion thus found is used as a constraint in the fit to the
June 1998 K$'$-band data.  Likewise, the values of $A_{\nu}$ and
$B_{\nu}$ may be assumed to be constant in time, and therefore only
the new value for the companion offsets need to be found from the more
limited data set of September 1998.  This procedure was iterated,
starting with every solution that provided an adequate fit to the June
1998 H-band data set, and converged to the best fit solution
summarized in Table~\ref{bintab3} and plotted in
Figure~\ref{mwc361fig}. Note that the best solution is found for
slightly different companion offsets at each epoch. However, the
difference is at the $2\sigma$ level, and therefore only of marginal
significance.

By substituting the values for $A_{\nu}$ and $B_{\nu}$ in
Equations~\ref{A} and \ref{B}, and given that the total fluxes in the
system are known (6.0 Jy at H and 8.1 Jy at K, see
Table~\ref{excesstable}), we have a system of 6 equations and 8
unknowns, and therefore the full solution for the individual fluxes
and visibilities is indeterminate.  However, in order to gain some
insight as to the amount of resolution on each component that is
implied, we solve the above system of equations under the two limiting
cases that the primary or secondary component is unresolved. The
results are summarized in Table~\ref{mwc361tab}. The table also shows
the Gaussian model FWHM ($\theta^G$) that corresponds to the
visibility determined for each component.  It can be seen from the
table that the implied sizes for the dust envelopes around each
component are comparable to those obtained for the other stars in our
sample under the Gaussian model of Section~\ref{gausssection}.

\citet{Corporon:1999} found some evidence of radial velocity
variations for the MWC~361 system in various absorption lines of He I
and Mg II, but did not have enough data to set a period. If it is the
star responsible for the radial velocity variations that we have
detected interferometrically, this system is a prime candidate for
continued observations, using both techniques, from which a full
orbital solution and individual stellar masses may be obtained.

\begin{deluxetable}{lccccccc}
\tablecaption{Binary solution for MWC~361-A. \label{bintab3}}
\tablecolumns{8}
\tablewidth{0pt}
\tablehead{
\colhead{Epoch \& Band} & \colhead{$A_{\nu}$} & \colhead{$B_{\nu}$} &
\colhead{$\xi_2$} & \colhead{$\eta_2$} & \colhead{N} & \colhead{Fit RMS} \\
& & & \colhead{(mas)} & \colhead{(mas)} & & \colhead{(\%)}
}
\startdata
June 1998 - H	 	& $0.772 \pm 0.005$ & $0.156 \pm 0.008$ & 
	$0.7 \pm 0.2$ & $18.2 \pm 0.2$ & 27 & 2.3 \\
June 1998 - K$'$ 	& $0.820 \pm 0.12\phn$ & $0.094 \pm 0.019$ & 
	\nodata\tablenotemark{a} & \nodata\tablenotemark{a} & 9 & 3.2 \\
September 1998 - H 	& \nodata\tablenotemark{a} & \nodata\tablenotemark{a} & 
	$2.4 \pm 1.1$ & $17.7 \pm 0.3$ & 13 & 6.3 \\
September 1998 - K$'$	& \nodata\tablenotemark{b} & \nodata\tablenotemark{b} &
	\nodata\tablenotemark{c} & \nodata\tablenotemark{c} & 5 & 2.8
\enddata
\tablenotetext{a}{Constrained to June 1998 - H solution}
\tablenotetext{b}{Constrained to June 1998 - K$'$ solution}
\tablenotetext{c}{Constrained to September 1998 - H solution}
\end{deluxetable}

\vspace{1in}

\begin{deluxetable}{cccccccc}
\tablecaption{MWC 361 limiting solutions. \label{mwc361tab}}
\tablecolumns{8}
\tablewidth{0pt}
\tablehead{
 \multicolumn{8}{c}{assume: $V_1^{H,K}=1.0$, then:}  \\ \cline{1-8}
\colhead{$F_1^H$} & \colhead{$F_2^H$} & \colhead{$V_2^H$} & 
\colhead{$\theta^G_2$} & \colhead{$F_1^K$} & \colhead{$F_2^K$} & 
\colhead{$V_2^K$} & \colhead{$\theta^G_2$}  \rule{0in}{3ex} \\
\colhead{(Jy)} & \colhead{(Jy)} & & \colhead{(mas)} & \colhead{(Jy)} & 
\colhead{(Jy)} & & \colhead{(mas)} 
}
\startdata
4.6 & 1.4 & 0.7 & 4.0 & 6.6 & 1.5 & 0.5 & 6.8 \\ \hline
\multicolumn{8}{c}{assume: $V_2^{H,K}=1.0$, then:}
\rule{0in}{3ex} \\ \cline{1-8}
$F_1^H$ & $F_2^H$ & $V_1^H$ & $\theta^G_1$ & $F_1^K$ & $F_2^K$ & $V_1^K$ & 
$\theta^G_1$ \rule{0in}{3ex} 
 \\[1ex] \hline
5.06 & 0.9 & 0.9 & 1.9 & 7.3 & 0.8 & 0.9 & 2.7
\enddata
\end{deluxetable}

\clearpage

\section{DISCUSSION} \label{conclusions}

The results of our study have demonstrated that, contrary to previous
belief, most HAEBE stars are easily resolvable in the near-IR by the
current class of ground based interferometers. Thus, the question of
the precise distribution of CS material in these young stars is one
that can potentially be solved using this technique.

Ideally, one would like to find a single physical model which can
uniquely explain all the sources. However, we know that at least 1 of
the 15 sources is a binary system (MWC~361-A), and moreover,
independent evidence (spectroscopic and photometric) points to the
binary nature of many other HAEBE stars.  Therefore, although the
results of fitting the visibility data to binary models are not
significantly better than the fits to simpler 1-parameter models, our
data are certainly consistent with such an interpretation, and it is
likely that this will turn out to be the correct explanation for at
least some of the sources in our sample. That said, we now consider
whether the ensemble of observations can be understood in terms of a
single model of the near-IR excess.  We find that, although our data
are too limited in spatial frequency coverage to uniquely determine a
model, the results for the ensemble of sources have features that
suggest the interpretation of the emission as arising in spherical
envelopes around the stars.

The accretion disk model fails in most cases, and therefore may be
ruled out as a candidate for the physical explanation of the IR
excess.  The problem is that, in general, the near-IR emitting region
must be located too close to the star in order to reproduce the
observed SEDs, resulting in predictions for the visibilities that are
high compared to the observations.  This remains true even if the
disks are allowed to be inclined to the observer, and with the
exceptions of MWC~297 and two of the unresolved sources, T~Ori and
MWC~147-A, the model can only reproduce the observations in some cases
with extreme values of the inclination angle and specific values of
the disk position angle.  Moreover, due to the larger inner holes and
higher accretion rates implied, an additional difficulty arises with
these solutions in that they exacerbate the difficulties raised by
\citet{Hartmann:1993} concerning the physical nature of the opacity
holes.  We note that, alternatively to tilting the disks, the
visibility data may be matched by considering a much steeper
temperature law in the inner edge of the disk ($T \propto r^{-8}$,
similar to the ring model), so that the H and K$'$ emission arise
essentially from the same radius. Thus, if an accretion disk is the
explanation for the near-IR excess, its inner structure must be quite
different from the generally accepted range of power law models.

It is also interesting to compare our results with the mm-wave
aperture synthesis observations of \citet*{Mannings:1997} (hereafter
MS), who concluded that the mm-wave circumstellar emission from HAEBE
stars was consistent with accretion disks.  Of the sources in our
sample, three were also part of their study: AB~Aur (unresolved in
continuum emission, resolved in molecular line emission), MWC~863
(unresolved in continuum, undetected in molecular lines), and MWC~275
(resolved in both continuum and molecular emission). For the two
resolved sources, the authors found an elongated structure with
inclinations $76\arcdeg$ and $58\arcdeg$ and position angles
$79\arcdeg$ and $126\arcdeg$, for AB~Aur and MWC~275 respectively. The
presence of an accretion disk was inferred based on the derived disk
sizes and masses, which would result in extinctions toward the stars
greater than observed if the material were spherically distributed,
and on line velocity maps, which reveal velocity gradients consistent
with orbiting material in Keplerian rotation.  As we have seen
however, the analysis of AB~Aur using our accretion disk model
indicates that the observed symmetry is inconsistent with the
inclination derived by MS. For MWC~275, the accretion disk model could
fit the data well only for inclinations $\gtrsim\,85\arcdeg$, again
inconsistent with the result of MS. Therefore, if the mm-wave data do
imply circumstellar disk geometry, it appears that at least in these
two examples the inner regions ($<1\,AU$) do not follow a simple
extrapolation from the disk structure at hundreds of AU from the star.  

It is clear from our analysis that both the Gaussian and ring models
provide good fits to the data (the only two exceptions being V594~Cas
and MWC~297, for which the ring model fails to reproduce the measured
sizes and excess near-IR fluxes). Moreover, it is a striking feature
of our data set that, except for the obvious binary, none of the
sources shows any departure from a brightness distribution that
appears circularly symmetric on the sky. This statement is most
significant for sources which have observations spanning a range of
baseline position angles.  Although the rotation of the baseline is
essentially zero in some sources, a more typical value is $\sim
30\arcdeg$, with a maximum of $65\arcdeg$ in the greatest case.  Thus,
to be consistent with the ensemble of observations, models in which
the emitting dust is distributed symmetrically around the star are
favored. In the case of AB~Aur, for which the baseline rotation is
among the largest ($50\arcdeg$), our results are consistent with the
overall symmetry of circumstellar material at radii $\geq 1 \arcsec$
found in visible light Hubble Space Telescope coronographic
observations \citep{Grady:1999}.  For most sources however,
it is also clear that the indication of circular symmetry needs to be
established more firmly with additional data covering a larger baseline
position angle range.

Physically the observed circular symmetry would correspond to a
spherical distribution of dust, as has been proposed in models by
\citet{Berrilli:1992}, \citet{Hartmann:1993}, \citet{Miro:1997} and
\citet{Pezzuto:1997}.  This interpretation is supported by the
additional result that, in the context of the Gaussian model, the
measured sizes and near-IR fluxes require the emission to have finite
optical depths, consistent with the fact that the central stars need to be
optically visible when viewed through these envelopes.  The ring model
in this scenario could then interpreted to represent emission from a
thin spherical shell, as has been proposed for AB~Aur by
\citet{Butner:1994}.

\begin{figure}[htbp]
\begin{center}
\includegraphics[height=4in]{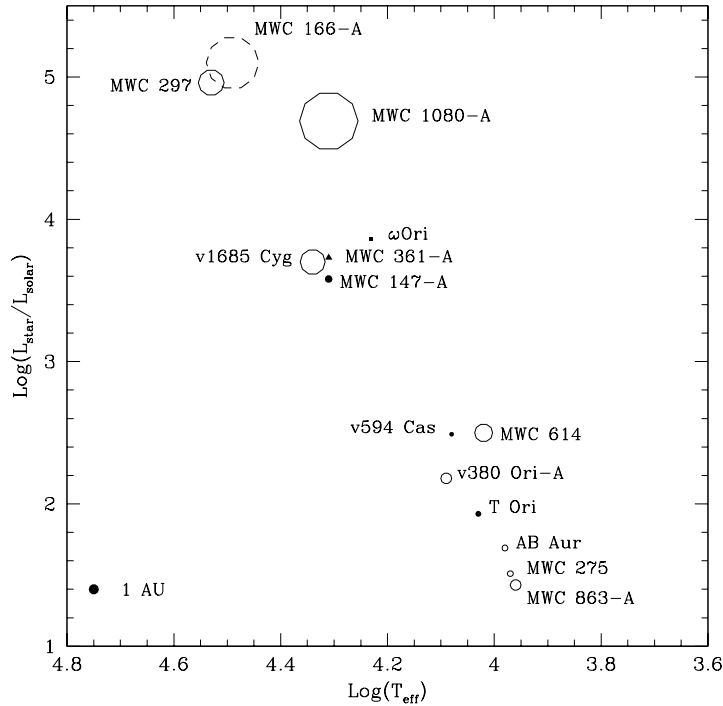}
\caption{Our sample in the HR diagram. The values plotted
correspond to the central photospheres, and the size of each symbol is
proportional to the measured linear size of the source of near-IR
excess.  Solid symbols proportional to the upper limits are used for
unresolved sources, and open symbols represent resolved sources (in
dashed lines if the size is a lower limit).  Different symbols are
used for $\omega$ Ori (square) because it has no IR excess and for
MWC~361 (triangle) because it is a binary system.
\label{hrfig}}
\end{center}
\end{figure}

With respect to the physical implications of this model, we find that
the properties of the excess are not strongly correlated with those of
the underlying star.  We show in Figure~\ref{hrfig} the location of
the stellar photospheres for the sources in our sample in the HR
diagram, based on our adopted effective temperatures
(Table~\ref{starfluxes}) and luminosities estimated as $L_{\star} = 4
\pi \, \sigma R_{\star}^2 \, T_{eff}^4$, which leads to similar values
as those in HSVK and \citet{Ancker:1998}. In the diagram, the sources
which were found to be resolved are shown using open symbols
proportional to the measured linear sizes; and solid symbols are used
for the size upper limits for unresolved objects. It can be seen that
although there is a tendency for the excess sources with largest sizes
and highest brightness to be located in the upper left part of the
diagram, the dependence of those two quantities with stellar
luminosity and effective temperature is rather weak.  We also find
that there are pairs of sources, (T~Ori,~V380~Ori-A) and
(MWC~147-A,~V1685~Cyg), in which the stars and IR excess are
essentially identical, but the sources of the IR excesses must differ
in size by more than a factor of two.  This might suggest that
different physical mechanisms are responsible for the near-IR emission
in these cases, and that there is no single phenomenon which scales
with the properties of the central star or the magnitude of the excess
in a simple way. Alternatively, if the same underlying mechanism is at
work in all cases, then it must have the property that the same IR
excess is produced by systems with very different physical scales.
Clearly, such conclusions place profound constraints on the nature and
physical distribution of the circumstellar material.

\acknowledgments

The authors wish to acknowledge M. G. Lacasse for invaluable help
during the observations. Instrument development and observations at
the IOTA are supported by funding from the SAO and the University of
Massachusetts. Work at the IOTA is also funded by NASA (NSG-7176 to
Harvard University and NAG5-4900 to SAO) and the NSF (ASR90-21181 to
IOTA consortium members and AST 95-28129 to the University of
Massachusetts).  RMG wishes to acknowledge that part of this work was
performed while he was a Michelson Postdoctoral Fellow, funded by
the Jet Propulsion Laboratory, which is managed for NASA by the
California Institute of Technology.

\end{document}